\documentclass[twocolumn, prd, aps,superscriptaddress, preprintnumbers, tightenlines, showpacs, nofootinbib, eqsecnum, amsfonts, amsmath, floatfix]{revtex4-1}

\usepackage{epsfig}
\usepackage{graphics}
\usepackage{graphicx}
\usepackage{amsmath}
\usepackage{amssymb}
\usepackage{timestamp}
\usepackage{bm}
\usepackage[usenames,dvipsnames,svgnames,table]{xcolor}
\usepackage{xspace}
\usepackage{wasysym}
\usepackage{times}
\usepackage{appendix}
\usepackage{lipsum}
\usepackage[nolist,nohyperlinks]{acronym}
\usepackage{feynmp}

\SetSymbolFont{letters}{bold}{OML}{cmm}{b}{it}
\SetSymbolFont{operators}{bold}{OT1}{cmr}{bx}{n}

\usepackage{float}

\makeatletter

    \def\CT@@do@color{%
      \global\let\CT@do@color\relax
            \@tempdima\wd\z@
            \advance\@tempdima\@tempdimb
            \advance\@tempdima\@tempdimc
    \advance\@tempdimb\tabcolsep
    \advance\@tempdimc\tabcolsep
    \advance\@tempdima2\tabcolsep
            \kern-\@tempdimb
            \leaders\vrule
                    \hskip\@tempdima\@plus  1fill
            \kern-\@tempdimc
            \hskip-\wd\z@ \@plus -1fill }
    \makeatother

\definecolor{ZurichBlue}{rgb}{.255,.41,.884} 		
\definecolor{ZurichRed}{rgb}{0.9, 0.1, 0} 			
\definecolor{ZurichGreen}{rgb}{.196,.504,.396} 		
\definecolor{ZurichYellow}{rgb}{1,.648,0} 			
\definecolor{dodgerblue}{rgb}{0.12, 0.56, 1.0}
\definecolor{azure}{rgb}{0.0, 0.5, 1.0}
\definecolor{awesome}{rgb}{1.0, 0.13, 0.32}
\definecolor{alizarincrimson}{rgb}{0.82, 0.1, 0.26}
\definecolor{mediumpurple}{rgb}{0.58, 0.44, 0.86}
\definecolor{lasallegreen}{rgb}{0.03, 0.47, 0.19}

\usepackage[backref=page,colorlinks=true,linkcolor=dodgerblue,citecolor=dodgerblue,urlcolor=dodgerblue]{hyperref}

\begin{document}

\newcommand{\be}{\begin{equation}}
\newcommand{\ee}{\end{equation}}
\newcommand{\ber}{\begin{eqnarray}}
\newcommand{\eer}{\end{eqnarray}}
\newcommand{\bea}{\begin{eqnarray}}
\newcommand{\eea}{\end{eqnarray}}
\newcommand{\ie}{i.e.}
\newcommand{\etal}{\textit{et al.}}
\newcommand{\rmi}{{\rm i}}
\newcommand{\bfr}{{\bf{r}}}
\newcommand{\bfF}{{\bf{F}}}
\newcommand{\bfk}{{\bf{k}}}
\newcommand{\bfx}{{\bf{x}}}
\newcommand{\bfq}{{\bf{q}}}
\newcommand{\bfd}{{\bf{d}}}
\newcommand{\bfC}{{\bf{C}}}
\newcommand{\bfW}{{\bf{W}}}
\newcommand{\bfmu}{{\bf{\mu}}}
\newcommand{\bftgr}{{\bf{theta}}^{\textrm{GR}}}
\newcommand{\bftmg}{{\bf{theta}}^{\textrm{MG}}}
\newcommand{\bft}{{\bf{\theta}}}
\newcommand{\calC}{{\mathcal{C}}}

\renewcommand{\Re}{\operatorname{Re}}
\renewcommand{\Im}{\operatorname{Im}}

\newcommand\GReq{\mathrel{\overset{\makebox[0pt]{\mbox{\normalfont\tiny\sffamily GR}}}{=}}}

\def\n{\noindent}

\newcommand{\Sussex}{Mathematical and Physical Sciences, University of Sussex, Falmer II, BN1 9QA, Brighton, United Kingdom}
\newcommand{\CNRS}{Institut de Physique Th\'eorique, Universit\'e Paris-Saclay, CEA, CNRS, F-91191, Gif sur Yvette, France}

\title{3D Weak Lensing: Modified Theories of Gravity}

\author{Geraint Pratten}
\affiliation{\Sussex}

\author{Dipak Munshi}
\affiliation{\Sussex}

\author{Patrick Valageas}
\affiliation{\CNRS}

\author{Philippe Brax}
\affiliation{\CNRS}

\begin{abstract}
Weak lensing (WL) promises to be a particularly sensitive probe of both the growth of large scale structure (LSS) as well as the fundamental relation between matter density perturbations and metric perturbations, thus providing a powerful tool with which we may constrain modified theories of gravity (MG) on cosmological scales. Future deep, wide-field WL surveys will provide an unprecedented opportunity to constrain deviations from General Relativity (GR). Employing a three-dimensional (3D) analysis based on the spherical Fourier-Bessel (sFB) expansion, we investigate the extent to which MG theories will be constrained by a typical 3D WL survey configuration including noise from the intrinsic ellipticity distribution $\sigma_{\epsilon}$ of source galaxies. Here we focus on two classes of screened theories of gravity: i) $f(R)$ chameleon models and ii) environmentally dependent dilaton models. We use one-loop perturbation theory combined with halo models in order to accurately model the evolution of matter power-spectrum with redshift in these theories. Using a Fisher information matrix based approach, we show that for an all-sky {\em spectroscopic} survey, the parameter $f_{R_0}$ can be constrained in the range $f_{R_0}< 5\times 10^{-6}(9\times 10^{-6})$ for $n=1(2)$ with a 3$\sigma$ confidence level. This can be achieved by using relatively low order angular harmonics $\ell<100$. Including higher order harmonics $\ell>100$ can further tighten the constraints, making them comparable to current solar-system constraints. We also employ a Principal Component Analysis (PCA) in order to study the parameter degeneracies in the MG parameters. Our results can trivially be extended to other MG theories, such as the K-mouflage models. The confusion from intrinsic ellipticity correlation and modification of the matter power-spectrum at small scale due to feedback mechanisms is briefly discussed.
\end{abstract}

\pacs{
}

\maketitle

\begin{acronym}
\acrodef{WL}[WL]{Weak Lensing}
\end{acronym}

\newcommand{\WL}{\ac{WL}\xspace}


\section{Introduction}
\label{sec:intro}
Numerous independent observations across a range of scales have firmly established the accelerated expansion of the Universe. This can be completely explained within General Relativity (GR) by the introduction of a finely tuned cosmological constant $\Lambda$ or by an additional smooth energy-momentum contribution known as \textit{dark energy} (DE). 
Alternatively, this could signal a deviation from GR on cosmological scales, so called
\textit{modified theories of gravity} (MG) \cite{Clifton11,Joyce14}.
Some of the simplest theories we could consider are that of a single scalar field with a sufficiently flat potential that provides the potential energy needed to drive an accelerated expansion. Such a scalar field could arise as a new form of matter or, as considered in this paper, as an additional scalar degree of freedom in the gravitational sector corresponding to some modification of GR. {\color{black}{These scenarios are generically plagued with problems ranging from an incomplete understanding of the role of quantum corrections to fine-tuning issues. For example, we often require that both the vacuum energy and mass of the scalar must be exceptionally small.}} The smallness of the vacuum energy constitutes nothing more than a reformulation of the cosmological constant problem whereas the ultra-light mass of the field posits that there should exist a new fifth-force at very large scales. {\color{black}{Such fifth forces are strongly constrained by solar system observations to the extent that we typically require some form of screening mechanism that suppresses the fifth force on these scales \cite{Vainshtein72,Damour94,Khoury03,Brax13a}.}} These \textit{screening mechanisms} schematically arise by introducing some mechanism that changes the nonlinear behaviour of the field at small scales whilst leaving the scalar field to be ultra light on linear cosmological scales. 

Fundamentally, the two approaches of (GR+DE) and MG are very different. However, there is often sufficient freedom in both of these approaches that they may be tuned to match any expansion history of the Universe. For the $f(R)$ and dilaton theories considered in this paper, the background dynamics will be the same as in GR \cite{Brax10,Brax12,Brax12a,Brax13}. However, in some models, such as the K-mouflage theories considered in \cite{Brax14,Brax15}, this is not true and the background dynamics can deviate from that of GR. On the other hand, the perturbative regime often breaks degeneracies between MG and DE. It is therefore instructive to consider observables that probe the evolution of perturbations in screened theories of gravity, in our case weak lensing. 

Many different parameterisations for perturbations in modified theories of gravity exist in the literature. One of the simplest possibilities is to consider linear cosmological perturbations in the quasistatic limit, where $k/a \gg H$ and we neglect time derivatives of the fields. Under this assumption, we can introduce two functions $\nu (k,a)$ and $\gamma (k,a)$ that parameterise deviations from GR via the modified Poisson equation \cite{Caldwell07,Amendola08}
\begin{align}
- k^2 \Psi &= 4 \pi \left( 1 + \nu \right) \, {\cal G}_{\rm{N}} \, a^2 \, \delta \rho_{\rm{M}} ,
\end{align}
\n
and
\begin{align}
\Phi &= \left( 1 + \gamma \right) \Psi.
\end{align}
\n 
Here, $\gamma$ is commonly referred to as the slip, $\mathcal{G}_N$ is Newton's constant and $\Phi$ and $\Psi$ are the metric potentials in the Newtonian gauge 
\begin{align}
ds^2 &= a^2 (\tau) \left[ - \left( 1 + 2 \Phi \right) d \tau^2 + \left( 1 - 2 \Psi \right) d \bfx^2 \right] ,
\label{Newtonian-gauge}
\end{align}
\n
where $\tau$ is the conformal time, $a(\tau)$ is the scale factor and $\bfx$ comoving coordinates. 

This parameterisation is convenient for phenomenological constraints from large scale structure as $\nu$ and $\gamma$ are very general functions of $k$ and $\tau$. It can be shown that this parameterisation works well in the linear regime but differs significantly from the theoretical predications in the non-linear regime, where screening effects will be important. The main reason why the $\nu-\gamma$ parameterisation fails, is that it does not correctly capture the environmental dependence of the screening mechanisms. The $\nu-\gamma$ type parameterisation will therefore not be valid in regimes ranging from mildly non-linear scales to solar system scales. Another alternative parameterisation that we could adopt is that based on effective field theories for large scale structure,  \cite{Creminelli08,Gubitosi12,Bloomfield12,Hu13,Hu16}. This formalism constructs an action in the Jordan frame, adopting the unitary gauge, such that the operators are invariant under time dependent spatial diffeomorphisms. At quadratic order, this yields 9 free functions that may be tuned to fit a given modified theory of gravity. In this paper, however, we will adopt the tomographic parameterisation of \cite{Brax11} which aims to cover a broad class of theories that exhibit a fifth force mediated by additional scalar degrees of freedom. The success of these theories implicitly relies upon screening mechanisms to suppress the fifth force in local, high-density environments. The tomographic approach can be shown to only require the temporal dependence of the mass $m(a)$ and the coupling to matter $\beta (a)$. Once these two variables are determined, then the behaviour of the theory in different regimes will be completely fixed \cite{Brax11}. 

As well as developments on the theoretical aspects of parameterising MGs, many different observational signatures have been proposed, such as galaxy clustering \cite{Pogosian08,Munshi15}, CMB weak lensing \cite{PlanckMG,Munshi16}, integrated Sachs-Wolfe (ISW) effect in the CMB \cite{Zhang06,Zhao10,PlanckMG}, galaxy-ISW cross-correlations \cite{Song07,Zhao10}, cluster abundances \cite{Jain08}, galaxy-clustering ratios \cite{Bel15}, redshift space distortions \cite{Jennings12}, weak-lensing \cite{Heavens07,Schmidt08,Harnois-Deraps15}, 21cm observations \cite{Brax12b,Hall13}, matter bispectrum \cite{Bernardeau11,GilMarin11} and many others. The typical scales modified by the theories of gravity considered here correspond to sub-horizon scales, meaning that the most stringent constraints come from large scale structure data sets. For $f(R)$ gravity, for example, one can constrain the model parameters to $|f_R| \leq 6.5 \times 10^{-5}$ at $95$\% confidence limit when taking a joint analysis of the CMB temperature power spectrum, the galaxy power spectrum and the baryon acoustic oscillation measurements. Taking the clustering ratio constraints inferred from the galaxy power spectrum, one finds that $|f_R| \leq 4.6 \times 10^{-5}$ at $95$\% confidence limit \cite{Bel15}. Similarly, it was shown that for a near-future 3D galaxy clustering analysis we should be able to tighten these constraints to $|f_R| \leq 2 \times 10^{-5}$ at a $95$\% confidence limit \cite{Munshi15}.

With several deep, wide-field galaxy surveys in the planning stages or underway, such as DES\footnote{http://www.darkenergysurvey.org/}, Euclid\footnote{http://www.euclid-ec.org/}, LSST\footnote{http://www.lsst.org/}, KiDS\footnote{http://kids.strw.leidenuniv.nl/overview.php} or CFHTLenS\footnote{http://www.cfhtlens.org/}, it is anticipated that galaxy clustering counts and weak lensing observations will be measured to an unprecedented level of accuracy. The relation between matter density perturbations and metric perturbations should be a particularly sensitive probe for constraining modified theories of gravity. Large scale structure probes, such as galaxy clustering, will offer high precision, small-scale constraints but will be strongly affected by non-linearities, galaxy biasing and other baryonic physics \cite{Munshi15}. This makes such galaxy clustering constraints sensitive to the detailed modelling of structure formation. Alternatively, we can consider the weak lensing of source galaxies induced by fluctuations in the gravitational potential along the line of sight, leading to observable distortions in the observed images \cite{Munshi06}. The deflection of photons by intervening structure will be the same in any metric theory of gravity meaning that the deflection will be largely unaltered in scalar-tensor theories. However, as scalar-tensor theories generically induce corrections to the Newtonian potential, there will be changes in the acceleration and clustering of galaxies. In the context of weak lensing, this simply means that the effect of the modified theories of gravity considered in this paper is to alter the matter power spectrum, growth function and gravitational potential.

Early weak lensing observations typically adopted a 2D flat-sky approach as the surveys only covered a small portion of the sky and lacked redshift information \cite{Hu98,Bacon00}. The next step was to fold in redshift information by performing a tomographic analysis, in which the data is binned into redshift slices \cite{Hu99,Hu02,Takada03}. This allows us to calculate the auto- and cross-correlations between redshift slices to give pseudo-3D results, meaning that less information is discarded. Instead, we focus on an all sky 3D formalism that implicitly takes into account photometric redshift information \cite{Heavens03}. This formalism is known as the spherical Fourier-Bessel (sFB) formalism and, at a statistical level, includes extra information that may be used to place tighter constraints on model parameters. This method has been recently applied to the CFHTLenS survey covering a 154 square degrees patch of sky with a median redshift of $z \sim 0.7$ and approximately 11 galaxies per square arcminute suitable for weak lensing \cite{Kitching14}.

In this paper we will study the constraints that may be placed on these screened modified theories of gravity using 3D weak lensing observations. We will focus on two classes of modified theories of gravity: i) $f(R)$ chameleon models and ii) environmentally dependent dilatons. Each of these models invokes a different mechanism for screening gravity but both can be described via the tomographic approach of \cite{Brax11}. These models lead to an enhancement of structure formation on quasi-linear and non-linear scales, i.e. $k \sim \lbrace 0.2 , 20 \rbrace h^{-1} \rm{Mpc}$, which will be within the reach of upcoming 3D weak lensing surveys. On non-linear scales, other observables, such as galaxy clustering, can give rise to large systematics due to the inherent uncertainty in galaxy bias and baryonic feedback mechanisms. These are partially circumvented by the cleaner nature of weak lensing.

The fundamental approach that we take here is similar to that of \cite{Harnois-Deraps15} in which the $\Lambda$CDM parameters are assumed to be determined to the percent level by Planck \cite{PlanckCosmology15}, while our modified theories of gravity lead to enhancements of up to 20 percent on small scales. This justifies our assumption that there is a fixed $\Lambda$CDM background with any uncertainty in the cosmology being treated as a systematic uncertainty. A more complete analysis would involve a full exploration of the parameter space using MCMC or Fisher matrix methods, we leave such a study to a future investigation. For now, we use a range of statistical tools to understand the constraints that may be placed on the screened theories of gravity in prototypical future 3D weak lensing surveys.

In Section \ref{sec:2} we introduce screened theories of gravity and detail the model parameters and the values taken in our study. In Section \ref{sec:3} we briefly discuss the formalism used to generate the non-linear matter power spectrum, relegating some technical details to Appendix \ref{app:nlmp}. We then introduce the spherical Fourier-Bessel formalism in Section \ref{sec:4} and discuss the machinery needed to describe 3D weak lensing observables, impact of modified theories of gravity, noise contributions and systematics. The bulk of the statistical analysis is presented in Section \ref{sec:5} in which we use a $\chi^2$ analysis to constrain the model parameters, a Fisher matrix analysis to estimate the $1\sigma$ errors on the parameters and a principal component analysis to determine the variance of the eigenvalues and which linear combination of eigenvalues may be constrained to the greatest degree. Section \ref{sec:6} presents a summary of the key results and some discussion on future topics of interest. 



\section{Modified Theories of Gravity}
\label{sec:2}
\subsection{Screening Mechanisms}
Modified theories of gravity are subject to many stringent constraints, both theoretical and observational. The scalar-tensor theories considered in this paper all aim to introduce infra-red (IR) modifications that may be able to resolve the dark energy problem and at least show some interesting effects on cosmological scales. Many of these scalar field models will have a coupling to the matter density that generally leads to a fifth force that is dependent on the gradient of the scalar field in dense regions. Such fifth forces are very tightly constrained by solar system measurements, such as the Cassini probe \cite{Bertotti03} and the Lunar Ranging experiment \cite{Williams12}. As a result of these observations, the fifth force must be highly suppressed necessitating some form of screening mechanism in order to ensure compatibility of our modified theory of gravity with the solar system constraints. 

Screening mechanisms come in three main flavours based on how the screening mechanism is implemented. Linearising about a field perturbation $\varphi = \bar{\varphi} + \delta \varphi$ in the presence of matter, the Lagrangian can be written in the Einstein-frame as 
\begin{align}
\mathcal{L} &\supset \underbrace{- \frac{1}{2} Z (\bar{\varphi}) \left( \partial \delta \varphi \right)^2}_{\textcolor{alizarincrimson}{\textrm{Vainshtein / K-mouflage}}} - \overbrace{\frac{1}{2} m^2_{\textrm{eff}} (\bar{\varphi}) \, \left( \delta \varphi \right)^2}^{\textcolor{lasallegreen}{\textrm{Chameleon}}} + \underbrace{\beta (\bar{\varphi}) \frac{\delta \varphi}{M_{\textrm{Pl}}} \delta T}_{\textcolor{dodgerblue}{\textrm{Damour-Polyakov}}} + \dots ,
\end{align}
\n
where $Z(\bar{\varphi})$ is the wavefunction normalisation or kinetic term, $m_{\textrm{eff}} (\bar{\varphi})$ is the effective mass and $\beta (\bar{\varphi})$ is the coupling to the trace of the energy-momentum tensor. The first class of screening mechanisms rely upon non-linearities in the kinetic term such that it becomes sufficiently large in dense environments. In this case, the fifth force constraints will be negligible due to a suppression of the effective coupling to matter {\textcolor{black}{$\beta (\bar{\varphi}) / \sqrt{Z(\bar{\varphi})} \ll 1$}}. This method is used by both the Vainshtein mechanism \cite{Vainshtein72} as well as the K-mouflage mechanism \cite{Babichev09,Brax14,Barreira15,Brax15}. The next class of screening mechanisms modify the effective mass of the field such that the field is massive in dense environments but ultralight on cosmological scales. This means that fifth forces will be suppressed on solar system scales whilst allowing for modifications to GR on IR scales. This type of screening is prototypically used by the Chameleon mechanism \cite{Khoury03} where the mass of the field grows with the matter density yielding a Yukawa like suppression of fifth forces. Finally, the last screening mechanism aims to reduce the coupling of the field in dense environments. The Symmetron \cite{Hinterbichler10} model utilises this approach and has a light mass in all environments with a coupling of the form $\beta (\varphi) \propto \varphi$. The model is equipped with a $Z_2$ symmetry breaking potential that gives rise to a phase transition that drives $\varphi$ to zero in dense environments suppressing fifth forces. Another possibility is that the coupling $\beta (\varphi)$ is driven to zero via the Damour-Polyakov \cite{Damour94} mechanism in which the coupling function is minimised in dense environments.

\subsection{Dilaton Models}
\label{Dilaton-Models}

\subsubsection{Theory}
The dilaton models are based on a breed of scalar fields that emerge from all versions of string theory. In the low energy limit string theory yields classical GR along with a four-dimensional scalar partner of the spin-2 graviton, the \textit{dilaton} $\varphi$. The vacuum expectation value (VEV) of the dilaton determines the string coupling constant $g_s = e^{\varphi / 2 M_{\rm pl}}$. At tree-level, the dilaton is massless with a gravitational-strength coupling to matter, placing it in conflict with current constraints on violations of the equivalence principle. A possible way to avoid this is to invoke mechanisms by which the dilaton can acquire a mass $m_{\varphi} \geq 10^{-3} \textrm{eV}$ suppressing deviations from GR at distances beyond the millimeter scale. Alternatively,  Damour and Polyakov \cite{Damour94} proposed a mechanism that naturally allows for a massless dilaton that can be reconciled with current experimental constraints. The Damour-Polyakov mechanism invokes string-loop modifications of an effective low-energy action to show that the graviton-dilaton-matter system in a cosmological setting naturally drives the dilaton $\varphi$ to $\varphi_m$, where $\varphi_m$ extremises the coupling functions $B^{-1}_{i} (\varphi)$ of the theory. This mechanism allows us to fix the value of the massless dilaton such that it decouples from matter, this is the so-called \textit{least coupling principle}. Under the Damour-Polyakov scenario, the coupling vanishes for a finite value of the dilaton whilst retaining an exponentially runaway potential that allows the dilaton to be displaced from its minimum without a coupling to matter. However, this result only holds when the string and Planck scales are of the same order of magnitude. If the string scale is lower than the Planck scale by a few orders of magnitude then the Damour-Polyakov mechanism is only at work in high density regimes, allowing solar system constraints to be evaded. This particular scenario is the \textit{environmentally dependent scenario} \cite{Olive07,Brax08}. 

In this paper we focus on the class of environmentally dependent dilaton models equipped with the Damour-Polyakov mechanism such that the coupling between the scalar field $\varphi$ and matter is driven to zero in dense environments. The scalar field remains light everywhere and thereby mediates a long-ranged screened force. The action describing this system in the Einstein frame has the following general scalar-tensor form
\begin{widetext}
\begin{align}
S&=\int d^4x \sqrt{-g} \left[ \frac{M_{\rm Pl}^2}{2} R - \frac{1}{2} (\nabla\varphi)^2 
- V(\varphi) - \Lambda_0^4 \right] + \int d^4x \sqrt{-\tilde{g}} \tilde{\cal L}_m( \psi^{(i)}_m,\tilde{g}_{\mu\nu}) ,
\label{eq:S-dilaton-def}
\end{align}
\end{widetext}
\n
where $M_{\textrm{pl}} = ( 8 \pi \mathcal{G}_{\rm N} )^{-1/2}$ is the reduced Planck mass in natural units, $g$ is the determinant of the metric $g_{\mu \nu}$ in the Einstein frame, $\tilde{g}$ is the determinant of the metric $\tilde{g}_{\mu \nu}$ in the Jordan-frame, $R$ is the Ricci scalar, $V(\varphi)$ the potential for a given theory and $\Lambda_0^4$ a cosmological constant contribution. The two frames are related by a conformal transformation
\begin{align}
\tilde{g}_{\mu \nu} &= A^2 (\varphi) g_{\mu \nu} .
\label{eq:conf}
\end{align}
\n
The matter field $\psi_m^{(i)}$ are governed by the Jordan-frame Lagrangian density $\tilde{\mathcal{L}}_m$ and the scalar-field $\varphi$ by the Einstein-frame Lagrangian density
\begin{align}
\mathcal{L}_{\varphi}  &= - \frac{1}{2} \left( \nabla \varphi \right)^2 - V (\varphi ) .
\end{align}
\n
The Klein-Gordon equation for the scalar field is modified due to the coupling of the scalar-field to matter
\begin{align}
\Box_g \varphi &= - \beta T + \frac{d V}{d \varphi} ,
\end{align}
\n
where $T$ is the trace of the energy-momentum tensor and the coupling of $\varphi$ to matter is defined by 
\begin{align}
\beta &= M_{\textrm{pl}} \, \frac{d \ln A}{d \varphi} .
\end{align}
\n
There is no explicit coupling between the scalar and the matter fields. The fifth force effects arise from the conformal transformation in Eq. (\ref{eq:conf}) via the gradients of $A$. For matter particles of mass $m$, the fifth force is given by ${\bf{F}} = - m c^2 \, \nabla \ln A$ \cite{Khoury03}. This may be written as an additional contribution $\Psi_{\rm{A}}$ to the Newtonian potential $\Psi_{\rm{N}}$
\begin{align}
\nabla^2 \Psi_{\rm{N}} &= 4 \pi \mathcal{G}_{\rm N} a^2 \delta \rho = \frac{3 \Omega_{M} H^2_0 \delta}{2 a} ,
\label{Psi-N-def}
\end{align}
\n
where the additional contribution is of the form
\begin{align}
\Psi_{\rm{A}} &= c^2 \left( A - \bar{A} \right) .
\end{align}
\n
We have assumed that $A (\varphi) \simeq 1$, as per experimental constraints on the variation of fermion masses. 

We have included the cosmological constant term $\Lambda^4_0$ in the Lagrangian in Eq. (\ref{eq:S-dilaton-def}) such that the minimum of the potential $V (\varphi)$ is zero for $\varphi \rightarrow \infty$. Alternatively, we could choose to interpret this as a non-zero minimum for the scalar-field potential. 

In the original tomographic dilaton models \cite{Brax10}, the scalar field potential $V(\varphi)$ and coupling $A (\varphi)$ had the following functional form
\begin{align}
V (\varphi) &\simeq V_0 \, \exp \left( - \frac{\varphi}{M_{\textrm{pl}}} \right) , \\
A (\varphi) &\simeq 1 + \frac{A_2}{2} \frac{\varphi^2}{M^2_{\textrm{pl}}} ,
\end{align}
\n
with $\lbrace V_0 , A_2 \rbrace$ being the free parameters of the theory. As can be seen, for $\varphi \sim 0$ the coupling to matter becomes negligible and the theory converges to GR. This theory can be generalised by retaining the form of the coupling function given above but generalising the potential $V(\varphi)$. We focus on models for which the effective potential 
\begin{align}
V_{\rm{eff}} (\varphi) &= V(\varphi) + \left[ A(\varphi) - 1 \right] \bar{\rho},
\label{eqn:Veff}
\end{align}
\n
has a minimum $\varphi (a)$ that depends on the scale factor due to the time variation of the matter density. This allows us to define the scalar mass at the minimum of the effective potential 
\begin{align}
m^2 &= \left. \frac{\partial^2 V_{\rm{eff}} (\varphi)}{\partial \varphi^2} \right|_{\varphi_m} .
\end{align}
\n
For models where $m^2 \gg H^2 / c^2$, the effective potential will be stable or quasistable and the dynamics will be completely determined by the minimum equation \cite{Brax12}
\begin{align}
\left. \frac{d V}{d \varphi} \right|_{\varphi_m} &= - \beta \, A \, \frac{\rho_m}{M_{\textrm{pl}}} .
\label{eqn:mineqn}
\end{align}
\n
Knowledge of the time evolution of the mass $m(a)$ and coupling $\beta(a)$ is sufficient in order to determine the bare potential $V(\varphi)$ and the coupling $A (\varphi)$. This tomographic reconstruction procedure allows us to define a one-to-one correspondence between the scale factor $a$ and the value of the field $\varphi (a)$. Given that $a$ is determined by $\rho_m$ this also defines a mapping from $\rho_m$ to $\varphi (\rho_m)$ using only the time evolution of $m(a)$ and $\beta (a)$ \cite{Brax11,Brax12}. Given the evolutions of these two variables, one can completely reconstruct the dynamics of the scalar field for densities ranging from cosmological scales down to solar system scales. 

\subsubsection{Derived Functions and Tomography}
Adopting the approach of \cite{Brax10,Brax11,Brax12,Brax12a,Brax13}, we can perturbatively expand in powers of $\delta \rho$ and $\delta \varphi$ with respect to the uniform background $( \bar{\rho} , \bar{\varphi})$. We can perform an expansion of the potential $V(\varphi)$ and coupling function $A(\varphi)$ such that 
\begin{align}
\label{eqn:beta_n_eqn}
n \geq 1 : \; \beta_n (a) &\equiv \beta \left[ \bar{\varphi} (a) \right] = M^n_{\textrm{pl}} \, \frac{d^n \ln A}{d \varphi^n} ( \bar{\varphi}) , \\
m^2 (a) &\equiv m^2 \left[ \bar{\varphi (a)}  , \bar{\rho} (a) \right] \\ & \nonumber = \frac{1}{c^2} \, \left[ \frac{d^2 V}{d \varphi^2} ( \bar{\varphi}) + \bar{\rho} \frac{d^2 A}{d \varphi^2} ( \bar{\varphi} ) \right] . 
\end{align}
\n
In addition it will be useful to define derivatives of the effective potential
\begin{align}
\label{eqn:kappa_n_eqn}
n \geq 2 : \, \kappa_n (\bar{\varphi} , \bar{\rho}) &= \frac{M^{n-2}_{\textrm{pl}}}{c^2} \, \frac{\partial^n V_{\textrm{eff}}}{d \varphi^n} (\bar{\varphi}) \\
&= \nonumber \frac{M^{n-2}_{\textrm{pl}}}{c^2} \, \left[ \frac{d^n V}{d  \varphi^n} (\bar{\varphi}) + \bar{\rho} \frac{d^n A}{d \varphi^n} (\bar{\varphi}) \right] .
\end{align}
\n
where $V_{\textrm{eff}}$ is defined by Eqn. (\ref{eqn:Veff}) and is the effective potential that enters the modified Klein-Gordon equation. We typically refer to these functions in terms of the scale factor $a(t)$ by defining $\beta_n (a) = \beta_n \left[ \bar{\varphi} (a) \right]$ and $\kappa_n (a) = \kappa_n \left[ \bar{\varphi} (a), \bar{\rho} (a) \right]$. As it is possible to reconstruct $V(\varphi)$ and $A(\varphi)$ through the two functions $\beta (a)$ and $m(a)$, a particular scalar-tensor model can then be defined by specifying the functional form for $\lbrace \beta (a) , m (a) \rbrace$. 

We adopt the parameterisation of \cite{Brax13} in which the coupling function is given by 
\begin{align}
A (\varphi) &= 1 + \frac{1}{2} \frac{A_2}{M^2_{\rm{pl}}} \varphi^2 ,
\end{align}
\n
and we specify the mass $m(a)$ instead of the potential $V(\varphi)$. The model is determined by the parameters $\lbrace m_0 , r , A_2 , \beta_0 \rbrace$ that set $m(a)$ and $\beta (a)$ by
\begin{align}
\label{eqn:param1}
m (a) &= m_0 \, a^{-r} , \\
\beta (a) &= \beta_0 \, \exp \, \left[ -s \frac{a^{2 r - 3} - 1}{3 - 2 r} \right] ,
\label{eqn:param2}
\end{align}
\n
where 
\begin{align}
s &= \frac{9 A_2 \, \Omega_{M} \, H^2_0}{c^2 m^2_0} .
\end{align}
\n
We can recover the original dilaton models of \cite{Brax10} by setting $r = 3/2$. We consider a series of 5 models $\lbrace A , B , C , D , E \rbrace$ that depend on the parameters of the theory $\lbrace s , \beta_0 , r , m_0, A_2 \rbrace$. The models $\lbrace A , B , C, E \rbrace$ study the dependence on $\lbrace s , \beta_0 , r , m_0 \rbrace$ respectively, keeping all other parameters fixed. In particular, the A-series systematically varies $s$, the B-series $\beta_0$, the C-series $r$ and the E-series $m_0$. The D-series jointly varies $\lbrace m_0 , s \rbrace$ such that it probes the dependence on $m_0$ with $A_2$ fixed. The parameters used for these models are explicitly given in Table \ref{tbl:dilaton}.

Together, these models allow us to study the phenomenological behaviour of the dilaton models as a function of the underlying model parameters. All models introduce deviations from GR at the level of 20 \% with respect to the matter power spectrum with the background cosmology being fixed to that of $\Lambda \rm{CDM}$. 

\subsubsection{Cosmological Dynamics}
By construction $A(\varphi) \simeq 1$ meaning that the Jordan and Einstein frame quantities will be nearly identical. In the adopted parameterisation $| \bar{A} - 1| \ll 1$ and $\bar{A} \simeq 1 + \beta^2 / (2 A_2)$, where $A_2 \sim (c m_0 / H_0)^2$. Solar system constraints place a lower limit on the mass of $m_0 \gtrsim 10^3 H_0 / c$ implying that $A_2 \gtrsim 10^6$ and hence 
\begin{align}
| \bar{A} - 1 | \lesssim 10^{-6} .
\end{align}
\n
Consequently, we will treat the Jordan frame and Einstein frame scale factor $\tilde{a} = \bar{A} a$, matter density $\tilde{\rho} = \bar{A}^{-4} \bar{\rho}$, cosmic time and expansion rates as being equal in the subsequent analysis. 

In the Einstein frame, the Friedmann equation has the form
\begin{align}
3 M^2_{\textrm{pl}} \, H^2 &= \bar{\rho} + \bar{\rho}_{\varphi} + \bar{\rho}_{\Lambda} .
\end{align}
\n
The background value of the scalar field is determined by the minimum, i.e. Eq. (\ref{eqn:mineqn}), and the evolution of the background scalar field and the background potential with respect to the scale factor is given by
\begin{align}
\frac{d \bar\varphi}{d a} &= \frac{3 \beta \bar{\rho} }{c^2 M_{\rm{pl}} \, a \, m^2 } ,\\
\frac{d \bar{V}}{d a} &= - \frac{3 \beta^2 \bar{\rho}^2}{c^2 M^2_{\rm{pl}} \, a \, m^2} .
\end{align} 
\n
For the models considered in this paper, the scalar field density is subdominant compared to the matter density, $\bar{\rho}_{\varphi} / \bar{\rho} \sim 10^{-6}$, and it is dominated by the potential term
\begin{align}
\frac{\dot{\bar{\varphi}}^2}{2 \bar{\rho}} \sim \left( \frac{H}{cm} \right)^4 \sim 10^{-12}, \quad \frac{\bar{V}}{\bar{\rho}} \sim \left( \frac{H}{cm} \right)^2 \sim 10^{-6}. \qquad 
\end{align}
Consequently, the Friedmann equation is governed by the matter density and cosmological constant ensuring that the background LCDM cosmological expansion, i.e. $3 M^2_{\textrm{pl}} H^2 = \bar{\rho} + \bar{\rho}_{\Lambda}$, is retained to an accuracy on order $10^{-6}$. 

In the quasistatic limit, the dynamics of the scalar field is given by the Klein-Gordon equation 
\begin{align}
\label{KG-Dilatons}
\frac{c^2}{a^2} \, \nabla^2 \varphi &= \frac{d V}{d \varphi} + \rho \frac{d A}{d \varphi} ,
\end{align}
\n
and the leading order perturbed Klein-Gordon equation reduces to \cite{Brax13}
\begin{align}
\frac{\delta \varphi}{M_{\rm{pl}}} &= - \frac{3 \beta \Omega_M a^2 H^2}{c^2 \left( a^2 m^2 + k^2 \right)} \delta .
\end{align}
\n
From this we find that \cite{Brax13,Harnois-Deraps15}
\begin{align}
\frac{\delta \rho_{\varphi}}{\delta \rho} \sim \left( \frac{H}{c m} \right)^2 \, \frac{1}{1 + k^2 / a^2 m^2} \lesssim 10^{-6} ,
\end{align}
i.e. fluctuations of the scalar field energy density are negligible in comparison with the matter density fluctuations. Coupled with the fact that $| \delta A | \lesssim 10^{-6}$, the modifications to the growth of structures do not arise from a different background cosmology or from perturbations to the scalar field energy density, but only from the fifth force acting on the matter density perturbations. Scalar field perturbations do not significantly alter the Einstein frame Newtonian potentials, such that $\Phi = \Psi = \Psi_{\rm N}$ to within an accuracy of $10^{-6}$ \cite{Munshi15}. The Newtonian potential obeys the standard Poisson equation as per Eq. (\ref{Psi-N-def}). However, unlike GR, we must add to the Newtonian potential a fifth-force potential $\Psi_A = c^2 \ln A$ that is not negligible and can lead to deviations in the matter power spectrum on the level of $10$ \% for the models considered in this paper. Whilst $| A - 1 | \leq 10^{-6}$ is negligible as compared to unity, it will not be negligible with respect to $| \Psi_{\rm N} | / c^2 \leq 10^{-5}$. 

{\renewcommand{\arraystretch}{1.3}%
\begin{table*}
\begin{center}
\begin{tabular}{| c | c | c | c | c | }
  \hline
  Model & $m_0 \;\; [h \, {\rm Mpc}^{-1}]$ & $r$ & $\beta_0$  & $s$ \\
  \hline
  \rowcolor[gray]{0.8} 
  (A1,A2,A3) & $(0.334,0.334,0.334)$ & $(1.00,1.00,1.00)$ & $(0.50,0.50,0.50)$ & $(0.60,0.24,0.12)$ \\
  \hline
  (B1,B3,B4) & $(0.334,0.334,0.334)$ & $(1.00,1.00,1.00)$ & $(0.25,0.75,1.00)$ & $(0.24,0.24,0.24)$ \\
  \hline
  \rowcolor[gray]{0.8} (C1,C3,C4) & $(0.334,0.334,0.334)$ & $(1.33,0.67,0.40)$ & $(0.50,0.50,0.50)$ & $(0.24,0.24,0.24)$ \\
  \hline
  (D1,D3,D4) & $(0.667,0.167,0,111)$ & $(1.00,1.00,1.00)$ & $(0.50,0.50,0.50)$ & $(0.06, 0.96, 2.16)$ \\
  \hline
 \rowcolor[gray]{0.8} (E1,E3,E4) & $(0.667,0.167,0.111)$ & $(1.00,1.00,1.00)$ & $(0.50,0.50,0.50)$ & $(0.24,0.24,0.24)$ \\
  \hline
\end{tabular}
\caption{Parameters describing the dilaton models considered in our study. The parameters are used to define the scalar potential
$V(\varphi)$ and the coupling function $A(\varphi)$ through the $\{\beta(a),m(a)\}$
parameterization.}
\label{tbl:dilaton}
\end{center}
\end{table*}

\subsection{$f(R)$ Models}
\label{fR-models}

\subsubsection{Theory}
A set of popular modifications to GR are the so-called fourth order theories of gravity (FOG). These arise as rather natural extensions to GR appearing in the low energy limit of various fundamental theories. In FOG, the Einstein-Hilbert action is modified by additional curvature functions that contain second derivatives of the metric. The resulting system of equations will be fourth order in nature
\begin{align}
\int d^4 x \, \sqrt{ - g} \, R \rightarrow \int d^4 x \, \sqrt{-g} \, f \left( R , R_{ab} R^{ab} , C_{abcd} C^{abcd} \right) ,
\end{align}
\n
where $R$ is the Ricci scalar, $R_{ab}$ the Ricci tensor and $C_{abcd}$ the Weyl tensor. Lovelock's theorem tells us that the field equations for a metric theory of modified gravity in a four-dimensional Riemannian manifold will admit higher than second order derivatives \cite{Lovelock71,Lovelock72}. These higher order terms will generically give rise to instabilities due to a theorem by Ostrogradski \cite{Ostrogradski}. The $f(R)$ models are a subclass of fourth-order theories of gravity that evade Ostrogradski instabilities due to the fact that they are degenerate. This just means that the highest derivative terms cannot be written as a function of the canonical variables. The resulting degrees of freedom can be completely fixed by a $g_{00}$ constraint, preventing ghost instabilities from arising in these theories. The $f(R)$ models were first introduced in \cite{Buchdahl70,Starobinsky80} and have subsequently been heavily discussed in the literature \cite{Sotiriou08,Clifton11}.

The action for the $f(R)$ theories considered here is given by
\begin{align}
S &= \int d^4 x \, \sqrt{-g} \, \left[ \frac{M^2_{\textrm{pl}}}{2} \left[ R + f(R) \right] - \Lambda^4_0 + \mathcal{L}_m ( \psi^{(i)}_m ) \right] ,
\end{align}
\n
where, again, we have explicitly included a cosmological constant term $\Lambda^4_0$. In this section, we will explicitly work in the Jordan frame with the Einstein frame metric being denoted by $g^E_{\mu \nu}$. We restrict ourselves to the high-curvature limit $f(R)$ theories that can be written in the following functional form \cite{Hu07,Brax11}
\begin{align}
f(R) &= - \frac{f_{R_0}}{n} \, \frac{R^{n+1}_0}{R^n}, \\
f_R &= \frac{d f(R)}{d R} = f_{R_0} \, \frac{R^{n+1}_0}{R^{n+1}} .
\end{align}
\n
The two free parameters in this theory are the normalisation $f_{R_0}$ and the exponent $n > 0$. These models are consistent with both solar system and Milky-Way constraints due to the chameleon mechanism for $| f_{R_0} | \leq 10^{-5}$. This particular class of models has been chosen to satisfy a number of desirable observational properties. Firstly, the cosmology must be consistent with LCDM at high redshifts due to CMB constraints. Secondly, there should be an accelerated expansion at low redshift with an expansion history that is sufficiently close to LCDM. Finally, the theory should reduce to GR as a limiting case. These constraints demand that 
\begin{align}
\lim_{R \rightarrow \infty} f(R) &= \textrm{const} , \\
\lim_{R \rightarrow 0} f(R) &= 0 .
\end{align}
\n
In such a theory, the background expansion will follow that of $\Lambda \rm{CDM}$  with the growth of structure deviating from GR on quasilinear and nonlinear scales. 

The class of $f(R)$ models can be shown to be equivalent to scalar-tensor theories expressed in the Einstein frame. For example, consider the conformal transformation $\tilde{g}_{\mu \nu} = A^{-2} (\varphi) g_{\mu \nu}$ where $A(\varphi) = \exp \left[ \beta \varphi / M_{\rm{pl}} \right]$ and $\beta = 1 / \sqrt{6}$. Under this transformation, the $f(R)$ theory is explicitly shown to correspond to an additional scalar degree of freedom $\varphi$ with a potential \cite{Brax11}
\begin{align}
V(\varphi) &= \frac{M^2_{\rm{pl}}}{2} \left( \frac{R f_R - f(R)}{(1 + f_R)^2} \right) ,
\end{align}
\n
where
\begin{align}
f_R &= \exp \left[ - \frac{2 \beta \varphi}{M_{\rm{pl}}} \right] - 1.
\end{align}
\n
This reformulation of the $f(R)$ theories is particularly elucidating in the sense that the screening mechanism clearly corresponds to the mass of the scalar field growing with the matter density with a Yukawa-like potential suppressing the fifth force in dense environments. So wherever the scalar field is small compared to the ambient Newtonian potential, screening will be efficient. 

Adopting the tomographic approach as before \cite{Brax11}, the $f(R)$ theories can be parameterised by the mass $m(a)$ and the coupling function $\beta (a)$ in terms of $a$  and the ambient background matter density $\bar{\rho} (a) = 3 \Omega_{M} H^2_0 M^2_{\rm{pl}} / a^3$. This establishes an explicit mapping from $\lbrace n , f_{R_0} \rbrace$ to $\lbrace m(a) , \beta (a) \rbrace$ 
\begin{align}
m(a) &= m_0 \, \left( \frac{4 \Omega_{\Lambda 0} + \Omega_{M} a^{-3}}{4 \Omega_{\Lambda 0} + \Omega_{M}}\right)^{(n+2)/2} , \\
m_0 &= \frac{H_0}{c} \, \sqrt{\frac{\Omega_{M} + 4 \Omega_{\Lambda 0}}{(n+1) |f_{R_0}|}} , \\
\beta (a) &= \frac{1}{\sqrt{6}} .
\end{align}
\n
Throughout this paper we adopt values of $n = \lbrace 1 , 2 \rbrace$ and $|f_{R_0}| = \lbrace 10^{-4} , 10^{-5} , 10^{-6} \rbrace$. Note that $|f_{R_0}| \sim 10^{-4}$ is currently ruled out by observations and will serve as a consistency check in our analysis. 

\subsubsection{Cosmological Dynamics}
The $f(R)$ models considered in this paper have been designed to closely mimic the background $\Lambda \rm{CDM}$ cosmology as $| f_R | \ll 1$. From the Friedmann equation we see that 
\begin{align}
3 M^2_{\rm{pl}} \left[ H^2 - \bar{f}_R \left(H^2 + \dot{H} \right) + \bar{f} / 6 + \bar{f}_{RR} H \dot{\bar{R}} \right] = \bar{\rho} + \bar{\rho}_{\Lambda} ,
\end{align}
\n
with dot derivatives denoting a derivative with respect to cosmic time $t$ and $f_{RR} = d^2 f / d R^2$. In the background, we see that $\bar{R} = 12 H^2 + 6 \dot{H}$. All the other terms are of order $| f_{R_0} | \, H^2$ such that the $\Lambda \rm{CDM}$ cosmology $3 M^2_{\rm{pl}} H^2 = \bar{\rho} + \bar{\rho}_{\Lambda}$ is recovered up to an accuracy of $10^{-4}$ for $|f_R| \leq 10^{-4}$. We can also check that the conformal factor $A(\varphi)$ is given by $A = (1 + f_R)^{-1/2}$ such that $| \bar{A} - 1 | \leq 10^{-4}$. This means that we can treat the background quantities in the Einstein and Jordan frames as being approximately equal and equal to the $\Lambda \rm{CDM}$ fiducial values to an accuracy of $10^{-4}$ or better. 

In terms of the Newtonian gravitational potential, $\Psi_{\rm N}$, the $f(R)$ theories introduce corrections to the Weyl scalars defined as in Eq.(\ref{Newtonian-gauge}) which, in the small-scale sub-horizon limit, reduce to \cite{Brax11}
\begin{align}
\label{eqn:fRPhi}
\Phi &= \Psi_{\rm N} - \frac{c^2}{2} \, \delta f_R ,  \\
\Psi &= \Psi_{\rm N} + \frac{c^2}{2} \, \delta f_R .
\label{eqn:fRPsi}
\end{align}
\n
Here $\delta f_R = f_R - \bar{f}_R$ and the subscript $\rm{N}$ denotes the Newtonian gravitational potential as defined in GR. These relations are calculated in the Jordan frame, therefore the modifications to gravity are directly imprinted in the metric potentials, unlike for the expressions obtained for the dilaton models considered previously that were derived in the Einstein frame. Conveniently, in the weak lensing potential the opposite sign contributions to the metric potentials (\ref{eqn:fRPhi})-(\ref{eqn:fRPsi}) will exactly cancel.

Finally, the fluctuations of the new scalar degree of freedom $\delta f_R$ are given by the Poisson equation
\begin{align}
\label{deltaR-deltarho}
3 \frac{c^2}{a^2} \nabla^2 \delta f_R &= \delta R - 8 \pi \mathcal{G}_{\rm{N}} \, \delta \rho ,
\end{align}
\n
and dynamics of matter particles will be given by solving the geodesic equation where the Newtonian metric potential is now replaced by the metric potentials in Eqns. (\ref{eqn:fRPhi}) and (\ref{eqn:fRPsi}).

\begin{figure}
\begin{center}
\includegraphics[width=90mm]{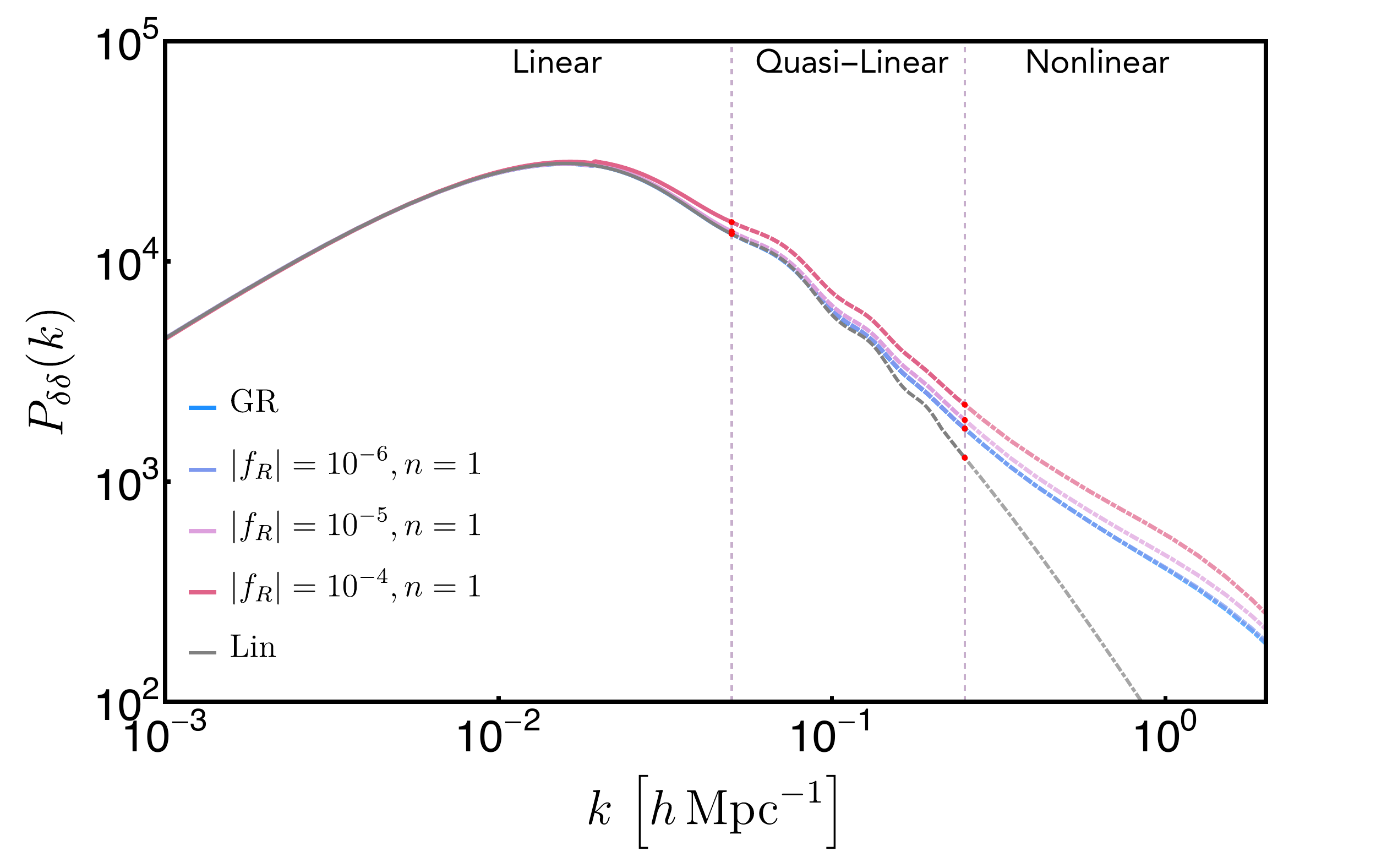}
\includegraphics[width=90mm]{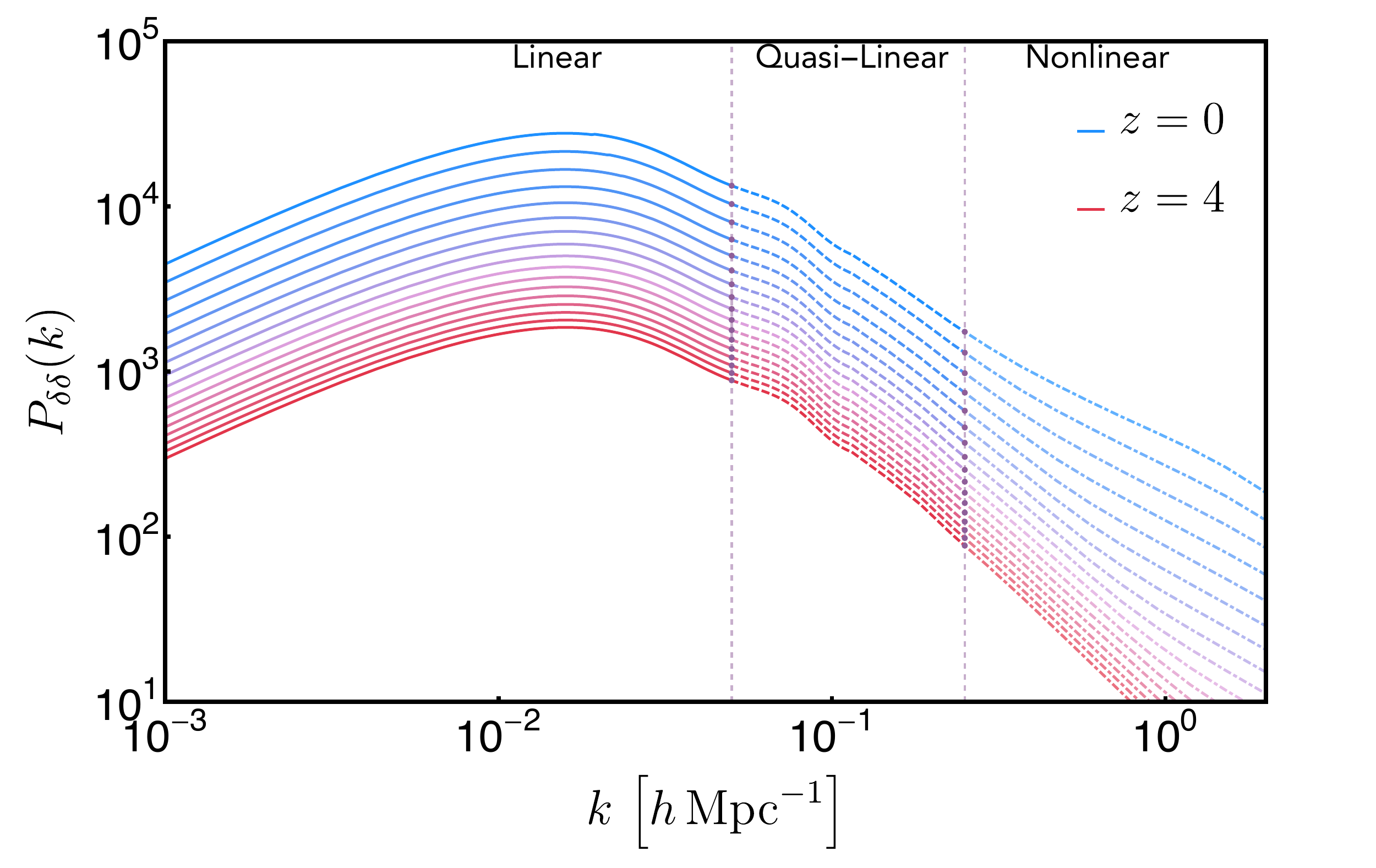}
\caption{
The nonlinear matter power spectrum $P_{\delta \delta} (k)$ for the $f(R)$ models (top) with $n=1$ and $|f_R| = \lbrace 10^{-4} , 10^{-5} , 10^{-6} \rbrace$. The bottom plot shows the redshift evolution of the nonlinear matter power spectrum in GR between $z = 0$ (top most line in lower plot) and $z = 4$ (bottom most line in lower plot). As can be seen in the upper plot, the modified theories of gravity lead to enhancement in structure formation on scales beyond $k \sim {\rm{few}} \times 10^{-2} \, h \, \rm{Mpc}^{-1}$ compared to GR. 
}
\label{fig:MPfR}
\end{center}
\end{figure}

{\renewcommand{\arraystretch}{1.3}%
\begin{table*}
\begin{center}
\begin{tabular}{| c | c | c | c | c | }
  \hline
  Model & $m_0 \;\; [h \, {\rm Mpc}^{-1}]$ & $r$ & $\beta_0$  & $s$ \\
  \hline
  \rowcolor[gray]{0.8} 
  $n = 1, \, | f_{R} | = \lbrace 10^{-4} , 10^{-5} , 10^{-6} \rbrace$  & $(0.042, 0.132, 0.417)$ & $(4.5,4.5,4.5)$ & $(0.408,0.408,0.408)$ & $(0.,0.,0.)$ \\
  \hline
  $n = 2, \, | f_{R} | = \lbrace 10^{-4} , 10^{-5} , 10^{-6} \rbrace$  & $(0.034, 0.108, 0.340)$ & $(6.0,6.0,6.0)$ & $(0.408,0.408,0.408)$ & $(0.,0.,0.)$ \\
  \hline
\end{tabular}
\caption{Parameters describing the $f(R)$ models considered in our study. These parameters are used to define the scalar potential
$V(\varphi)$ and the coupling function $A(\varphi)$ through the $\{\beta(a),m(a)\}$
parameterization.}
\label{tabular:tab2}
\end{center}
\end{table*}

\section{Non-Linear Power Spectra}
\label{sec:3}
The dilaton and $f(R)$ models reproduce the smooth background expansion history of $\Lambda \rm{CDM}$ cosmology to within a level of accuracy that cannot be detected by
observations. In order to study the effect of these modified theories of gravity we therefore need to move to the perturbative regime and study the evolution of the matter density and metric perturbations. At lowest order, modified theories of gravity typically result in a scale and time-dependent modification to the Newtonian gravitational constant ${\cal G}_{\rm N}$. In the quasilinear and nonlinear regime the modifications become much more sensitive to the particular screening mechanism which in turn depends nonlinearly on the environment. This results in modifications to the equations of motion, inducing modifications to the dynamical and statistical properties of matter density clustering. This can be seen in Figure \ref{fig:MPfR} where the mildly non-linear and non-linear scales show an enhancement in structure formation over that of GR. We also plot the change in the matter power spectrum as a function of redshift between $z=0$ and $z=4$. A detailed discussion of the prescription used to generate the nonlinear matter power spectra via the single stream approximation and halo modelling is given in Appendix \ref{app:nlmp}.

This approach is built on the techniques developed in \cite{Crocce05,Valageas13} for $\Lambda \rm{CDM}$ cosmologies and was subsequently extended to modified theories of gravity in \cite{Brax12a,Brax13,Brax14}. This approach combines results in 1-loop perturbation theory with results from halo models in order to extend the domain of validity of the nonlinear matter power spectrum to $k \sim 1 \, h \, \rm{Mpc}^{-1}$.

\section{3D Weak Lensing}
\label{sec:4}
\subsection{Weak Lensing Introduction}
\begin{figure*}[t!h]
\begin{center}
\includegraphics[width=110mm]{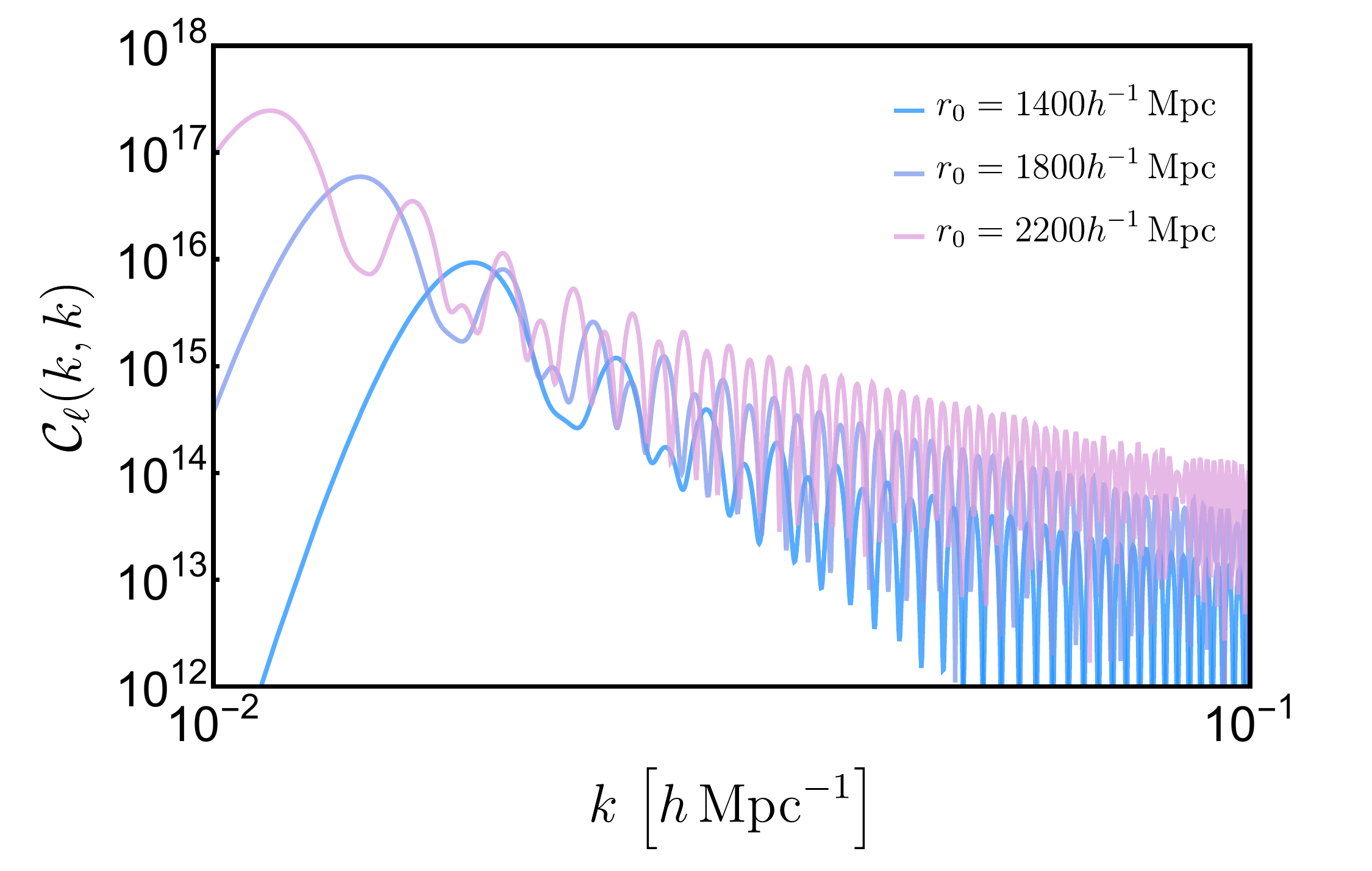}
\caption{
3D weak lensing lensing potential $C^{\phi \phi}_{\ell} (k_1 , k_1)$ in GR with $r_0 \in \lbrace1400, 1800, 2200 \rbrace \, h^{-1} \textrm{Mpc}$ and $\ell = 40$. Note that we have left $c = 1$ and neglected the amplitude pre-factor $A = 16 / \pi^2 c^4$, which is just some fixed scaling of the amplitude. 
}
\label{fig:3DWLGR}
\end{center}
\end{figure*}

Weak gravitational lensing is a particularly powerful probe of the epoch in cosmological history in which the transition from dark matter dominance to dark energy dominance occurs \cite{Munshi06}. This makes weak lensing observables particularly suitable for studies of DE or MG, where baryonic physics on small scales can be neglected circumventing the need for detailed modelling of various effects including bias between dark matter and baryonic matter. 

In the Newtonian gauge, scalar perturbations to the metric can be completely characterised by the two Weyl potentials, $\Phi$ and $\Psi$. If we take GR and neglect anisotropic stresses, an assumption which will be very reasonable on large scales
in $\Lambda \rm{CDM}$ and most generic smooth DE models, we find that 
$\Phi=\Psi= \Psi_{\rm{N}}$, where the Newtonian potential can be directly related to density perturbations via the Poisson equation (\ref{Psi-N-def}).
In many modified theories of gravity, these assumptions break down. This can be seen explicitly in the Jordan-frame expressions for the metric potentials in $f(R)$, Eqns. (\ref{eqn:fRPhi}) and (\ref{eqn:fRPsi}). 

However, as null geodesics are conformally invariant, the geodesics will only depend on the conformally-invariant part of the Riemann tensor, the Weyl part. This means that weak lensing is not sensitive to the individual metric potentials but instead to the linear combination of potentials given by
\begin{align}
\label{Phi-WL-def}
\Phi_{\textrm{WL}} &= \frac{\Phi + \Psi}{2} .
\end{align}
\n
Then, we find that for all the scenarios considered in this paper, the $\Lambda$CDM
cosmology, the Dilaton models and the $f(R)$ theories, we have
\begin{align}
\label{Phi-WL}
\Phi_{\textrm{WL}} &= \Psi_{\rm N} ,
\end{align}
\n
where $\Psi_{\rm N}$ is again the Newtonian potential defined by the standard
Poisson equation (\ref{Psi-N-def}).

\subsection{Spherical Fourier-Bessel Formalism}

Spherical coordinates will be a natural choice in the analysis of future cosmological data sets as, by an appropriate choice of coordinates, we can place the observer at the origin. Future surveys promise to yield both large (i.e. wide angle) and deep (i.e. large radial coverage) coverage of the sky, necessitating a simultaneous treatment of the extended radial coverage and spherical sky geometry. A natural basis for such an analysis is given by the spherical Fourier-Bessel (sFB) basis. In this section, we follow \citep{Heavens03, Castro05,Pratten13,Pratten14,Munshi15} and outline the conventions used for the sFB formalism in this paper. 

Let us consider a 3D random field $\zeta (r,\hat{\Omega})$ with $\hat{\Omega}$ denoting the angular coordinate on the surface of a sphere and $r$ denoting the comoving radial distance. In the 3D case, the eigenfunctions of the Laplacian will be constructed from products of the spherical Bessel functions of the first kind $j_{\ell} (kr)$ and spherical harmonics $Y_{\ell m} (\hat{\Omega})$ with eigenvalues of $-k^2$. For simplicity, we assume a flat background Universe and the sFB decomposition of the homogeneous 3D field reduces to
\begin{align}
{\zeta} (r,\hat{\Omega}) &= \int_0^{\infty}  dk \, \displaystyle\sum_{\lbrace \ell m \rbrace} \, \zeta_{\ell m} (k) \, Z_{k \ell m} (r,\theta,\varphi) , 
\end{align}
\n
where we have introduced the orthonormal sFB basis functions
\begin{align}
Z_{k \ell m} (r , \theta , \varphi ) &= \sqrt{\frac{2}{\pi}} \, k \, j_{\ell} (kr) \, Y_{\ell m} (\theta , \varphi) .
\end{align}
\n
The inverse relation is given by
\begin{align}
{\zeta}_{\ell m}(k) &= \int_0^{\infty} dr \, r^2 \, \int d\hat{\Omega} \,\zeta(r, \hat{\Omega}) \, Z_{k \ell m}^*(r,\theta,\varphi).
\label{eqn:zeta}
\end{align}
\n
This is something of a spherical analogue to the conventional Cartesian Fourier decomposition. 
In particular, defining the normalisation of the 3D Fourier transform and power spectrum as
\begin{align}
\zeta({\bf r}) &= \frac{1}{(2\pi)^{3/2}} \int d{\bf k} \, e^{{\rm i} {\bf k} \cdot {\bf r}} \, \zeta({\bf k}) , \\
\langle \zeta({\bf k}) \zeta^{\ast}({\bf k}') \rangle &= P_{\zeta \zeta }(k) \, \delta_D({\bf k} - {\bf k}') ,
\end{align}
the sFB coefficients and the Fourier modes can be related as
\begin{align}
\zeta_{\ell m}(k) &= {i}^{\ell} k \int d \hat{\Omega} \, Y_{\ell m}^{\ast}(\hat{\Omega}) \, \zeta (k,\hat{\Omega}) , \\
\zeta ({k,\hat{\Omega}}) &= \frac{1}{k} \displaystyle\sum_{\lbrace \ell m \rbrace} (-i)^{\ell} \zeta_{\ell m}(k) 
Y_{\ell m}(\hat{\Omega}) ,
\end{align}
\n
where we have introduced the conventional Dirac delta function $\delta_D (x)$. The sFB power spectrum can now be defined as
\begin{align}
\langle \zeta_{\ell m}(k) \zeta^{\ast}_{\ell^{\prime}m^{\prime}}(k^{\prime}) \rangle &= {\cal C}^{\zeta \zeta}_{\ell} (k) \, 
\delta_D(k - k^{\prime}) \, \delta_{\ell \ell^{\prime}} \, \delta_{m m^{\prime}} ,
\end{align}
\n
where we see that $ {\cal C}^{\zeta \zeta}_{\ell} (k) = P_{\zeta \zeta }(k)$. This result implicitly assumes all-sky coverage and neglectes any galactic cut-off, radial selection function or other discrete effects such as masking.

\subsection{The Lensing Potential in the sFB Formalism}
In gravitational weak lensing we can associate the weak lensing potential $\phi$ for some arbitrary source at comoving position $\bfr = (r,\theta,\varphi)$ to the underlying Weyl potential $\Phi_{\rm{WL}}$ via a line-of-sight integral 
\begin{align}
\phi (\bfr) &\equiv \phi (r , \hat{\Omega}) = \frac{2}{c^2} \, \int^r_0 d r^{\prime} \, F_K (r, r^{\prime} ) \, \Phi_{\textrm{WL}} (r^{\prime} , \hat{\Omega}) \\
F_K (r , r^{\prime}) &\equiv \frac{f_K (r - r^{\prime})}{f_K (r) \, f_K (r^{\prime})} .
\end{align}
\n
This expression implicitly uses the Born approximation in which the path of null geodesics is assumed to correspond to the unperturbed background path. Similarly, it also neglects any coupling between lenses along the line of sight. The lensing potential $\phi$ is conventionally treated as a 2D radial projection of the underlying 3D Weyl potential, e.g. \cite{Lewis06}. 

The weak-lensing potential can be harmonically decomposed as follows:
\begin{align}
\left[ \Phi_{\textrm{WL}} \right]_{\ell m} (k) &= \int d^3 \bfr \, \Phi_{\textrm{WL}} (\bfr) \, Z_{k \ell m}^*(\bfr) ,
\end{align}
\n
with an inverse relation
\begin{align}
\Phi_{\textrm{WL}} (\bfr) &= \displaystyle\sum_{\ell = 0}^{\infty} \, \displaystyle\sum_{m = - \ell}^{m = + \ell} \, \int dk \, \left[ \Phi_{\textrm{WL}} \right]_{\ell m} (k) \, Z_{k \ell m} (\bfr) .
\end{align}
 
Using the above decompositions, the harmonic decomposition of the lensing potential $\phi_{\ell m} (k)$ and the 3D gravitational potential $\left[ \Phi_{\textrm{WL}} \right]_{\ell m} (k, r)$ can be related by \cite{Castro05}
\begin{widetext}
\begin{align}
\phi_{\ell m} (k) &= \frac{4 k}{\pi c^2} \int^{\infty}_0 d k' \, k' \, \int^{\infty}_0  dr \, r^2 \, j_{\ell} (k r) \int^r_0 d r' \, F_K (r , r' ) \, j_{\ell} (k' r') \, \left[ \Phi_{\textrm{WL}} \right]_{\ell m} (k' ; r' ) .
\end{align}
\end{widetext}
\n
In GR and in the modified gravity models studied in this paper, the weak lensing potential $[\Phi_{\rm WL}]_{\ell m}(k;r)$ can be related to the matter overdensity $\delta_{\ell m} (k;r)$ via the usual Poisson relation (\ref{Psi-N-def}),
using the relation (\ref{Phi-WL}),
\begin{align}
\Phi_{\ell m} (k;r) & = - \frac{3}{2} \frac{\Omega_m H^2_0}{k^2 a(r)} \, \delta_{\ell m} (k;r)
\label{eq:GRPhi}
\end{align}
\n
These sets of equations explicitly determine the relationship between $\phi$, $\Phi_{\rm{WL}}$ and $\delta$ \cite{Castro05}. As the statistics of $\delta$ and $\Phi_{\rm{WL}}$ can be explicitly calculated, then 3D weak lensing through the lensing potential $\phi$ will allow us to probe the underlying cosmological or modified gravity model parameters. 
These expressions hold for the $\Lambda$CDM cosmology as well as for the Dilaton and
$f(R)$ models studied in this paper, as the background is not modified (within an accuracy of $10^{-4}$) and the weak lensing potential is still given by the standard Poisson equation
(\ref{Phi-WL}).

\subsection{3D Weak Lensing}
Most of the discussion has focused on the weak lensing potential $\phi$ whereas the actual lensing observables are the magnification $\kappa$, or the isotropic convergence scalar field, and the complex shear $\gamma (\bfr) = \gamma_1 (\bfr) + i \gamma_2 (\bfr)$, which corresponds to two orthogonal modes of distortion. Weak lensing shear is a spin-2 object and is fundamentally tensorial in nature, hence the appropriate basis harmonics will be spin-weighted spherical harmonics \cite{Stebbins96}. In this section we briefly review the full-sky 3D weak lensing formalism. This approach was first introduced in \cite{Heavens03} and subsequently generalised to a full tensorial form in \cite{Castro05}. Our treatment will follow the presentation of \cite{Castro05}. Note that weak gravitational lensing implicitly refers to the regime in which $| \gamma | \ll 1$ and $| \kappa | \ll 1$. 

The distortion of null geodesics on some 2D surface at a comoving distance of $\bfr$ induced by weak gravitational lensing via some intervening structure is given by 
\begin{align}
\left[ \nabla_i \nabla_j - \frac{1}{2} g_{ij} \nabla^2 \right] \, \phi (\bfr) &= \left[ \gamma_1 (\bfr) \sigma_3 + \gamma_2 (\bfr) \sigma_1 \right]_{ij} ,
\end{align}
\n
where $\sigma_i$ are the Pauli spin matrices, $g_{ij}$ is just the metric of the 2-sphere and $\nabla_i$ are covariant derivatives on the 2-sphere. Alternatively, we can study the isotropic convergence scalar field which is defined by the Laplacian of the lensing potential
\begin{align}
\left[ \kappa (\bfr) \right]_{ij} &= \kappa (\bfr) \, I_{ij} = \frac{1}{2} g_{ij} \, \nabla^2 \, \phi (\bfr) ,
\end{align}
\n
with $I_{ij}$ the identity matrix. In a spherical polar coordinate system $\lbrace \theta , \varphi \rbrace$, the weak lensing shear and convergence field tensor are explicitly given by \cite{Castro05}
\begin{widetext}
\begin{align}
\left[ \gamma (\bfr) \right]_{ij} &= \begin{bmatrix}
\frac{1}{2} \left[ \nabla_{\theta} \nabla_{\theta} - \csc^2 \theta \, \nabla_{\varphi} \nabla_{\varphi} \right] & \nabla_{\varphi} \nabla_{\theta} \\
\nabla_{\varphi} \nabla_{\theta} & \frac{1}{2} \left[ \nabla_{\varphi} \nabla_{\varphi} - \sin^2 \theta \, \nabla_{\theta} \nabla_{\theta} \right]
\end{bmatrix} \, \phi (\bfr) \\
\left[ \kappa (\bfr) \right]_{ij} &= \begin{bmatrix}
\frac{1}{2} \left[ \nabla_{\theta} \nabla_{\theta} + \csc^2 \theta \, \nabla_{\varphi} \nabla_{\varphi} \right] & 0 \\
0 & \frac{1}{2} \left[ \nabla_{\varphi} \nabla_{\varphi} + \sin^2 \theta \, \nabla_{\theta} \nabla_{\theta} \right]
\end{bmatrix} \, \phi (\bfr)
\end{align}
\end{widetext}
\n
The complex shear can now be split into two complex potentials $\phi_E$ and $\phi_B$ such that 
\begin{align}
\gamma (\bfr) &= \gamma_1 (\bfr) + i \gamma_2 (\bfr) = \frac{1}{2} \eth \eth \left[ \phi_E (\bfr) + i \phi_B (\bfr) \right] , \\
\gamma^{\ast} (\bfr) &= \gamma_1 (\bfr) - i \gamma_2 (\bfr) = \frac{1}{2} \bar{\eth} \bar{\eth} \left[ \phi_E (\bfr) - i \phi_B (\bfr) \right] ,
\end{align}
\n
where $\eth$ and $\bar{\eth}$ are differential operators that act as spin-raising and spin-lowering operators. The potentials $\phi_E$ and $\phi_B$ correspond to the electric (even parity) and magnetic (odd parity) parts of the field. Weak gravitational lensing alone is sourced by the real part of the field necessitating that $\phi_E = \phi (\bfr)$ and $\phi_B = 0$. As the shear field induced by weak lensing is of pure electric type, non-zero values for $\phi_B$ can be a robust diagnostic for systematics or foreground contamination.

The two real scalar potentials $\phi_E$ and $\phi_B$ will completely characterise the distortion field induced by weak lensing and can be expressed in the sFB basis as follows \cite{Castro05}
\begin{align}
\phi_E (\bfr) &= -2 \int^{\infty}_0 dk \, \displaystyle\sum_{\lbrace \ell m \rbrace} \, \sqrt{\frac{(\ell -2)!}{(\ell +2)!}} \, E_{\ell m} (k) \, Z_{k \ell m} (r,\theta,\varphi) ,\\
\phi_B (\bfr) &= -2 \int^{\infty}_0 dk \, \displaystyle\sum_{\lbrace \ell m \rbrace} \, \sqrt{\frac{(\ell -2)!}{(\ell +2)!}} \, B_{\ell m} (k) \, Z_{k \ell m} (r,\theta,\varphi) .
\end{align}
\n
The shear field $\gamma (\bfr)$ can be decomposed into a spin-weighted spherical Fourier-Bessel basis via
\begin{align}
\gamma (\bfr) &= \int^{\infty}_0 dk \, \displaystyle\sum_{\lbrace \ell m \rbrace} \, {_2}\gamma_{\ell m} (k) \, {_2}Z_{k \ell m} (r , \theta , \phi ) ,
\end{align}
\n
where we have introduced the spin weighted sFB basis functions
\begin{align}
_{s}Z_{k \ell m} (r , \theta , \varphi ) &= \sqrt{\frac{2}{\pi}} \, k \, j_{\ell} (kr) \, _{s}Y_{\ell m} (\theta , \varphi) .
\end{align}
\n
and the spin-weighted expansion coefficients can be related to the electric $E_{\ell m}$ and magnetic $B_{\ell m}$ harmonics via
\begin{align}
{_2}\gamma_{\ell m} (k) &= - \left[ E_{\ell m} + i \, B_{\ell m} \right] (k) , \\
{_{-2}}\gamma_{\ell m} (k) &= - \left[ E_{\ell m} - i \, B_{\ell m} \right] (k) .
\end{align}
\n
Given that for weak lensing we require that $\phi_B = 0$, then the complex shear field can be related to the underlying lensing potential $\phi (\bfr)$ as via \cite{Castro05}
\begin{align}
\gamma (\bfr) &= \frac{1}{2} \eth \eth \, \phi (\bfr) , \qquad \gamma^{\ast} (\bfr) = \frac{1}{2} \bar{\eth} \, \bar{\eth} \, \phi (\bfr) .
\end{align}
\n
By performing a sFB expansion of the lensing potential, and acting upon this with the spin-raising and spin-lowering operators, we can relate the $E_{\ell m}$'s and $B_{\ell m}$'s to the lensing potential harmonics
\begin{align}
E_{\ell m} (k) &= - \frac{1}{2} \sqrt{ \frac{( \ell + 2 )!}{( \ell - 2 )!} } \phi_{\ell m} (k) \qquad B_{\ell m} = 0 .
\end{align} 
\n
Deviations from $B_{\ell m} = 0$ should be a good discriminator of systematic effects. The 3D spin-weight 2 shear coefficients are related to the lensing harmonics via an $\ell$-weighted prefactor \cite{Castro05}
\begin{align}
\label{eqn:shear_lensing_relation}
{_2}\gamma_{\ell m} (k) &= _{-2}\gamma_{\ell m} (k) = \frac{1}{2} \sqrt{ \frac{( \ell + 2 )!}{( \ell - 2 )!} } \phi_{\ell m} (k) .
\end{align}
\n
This final expression will allow us to relate the shear lensing spectra $\calC^{\gamma \gamma}_{\ell} (k_1, k_2)$ to the lensing potential spectra $\calC^{\phi \phi}_{\ell} (k_1 , k_2)$.

\subsection{3D Weak Lensing Power Spectra}
\subsubsection{Lensing Potential and Lensing Shear Spectra}
The full 3D sFB decomposition of the Weyl potential $\Phi$ and the lensing potential $\phi$ can be used to define power spectra for via the harmonics in the usual way
\begin{align}
\langle \Phi_{\ell m} (k ; r) \, \Phi^{\ast}_{\ell^{\prime} m^{\prime}} (k^{\prime} ; r) \rangle &= \mathcal{C}^{\Phi \Phi}_{\ell} (k ; r) \, \delta_{\textrm{D}} (k - k^{\prime}) \, \delta^K_{\ell \ell^{\prime}} \, \delta^K_{m m^{\prime}} , \\
\langle \phi_{\ell m} (k) \, \phi_{\ell^{\prime} m^{\prime}} (k^{\prime}) \rangle &= \mathcal{C}^{\phi \phi}_{\ell} (k,k^{\prime}) \, \, \delta^K_{\ell \ell^{\prime}} \, \delta^K_{m m^{\prime}} ,
\end{align}
\n
where $\delta^K_{ab}$ are the Kronecker delta functions. 
Note that the lensing potential is not homogeneous and isotropic in 3D space but rather a 2D projection at a source of comoving distance $r$ of the underlying gravitational potential $\Phi$ between us and the source. This means that it will be homogeneous and isotropic on the 2-sphere but this will not hold in the radial direction \cite{Castro05}, hence the difference in the structure of the power spectra.

Using the harmonic decomposition of the lensing potential, the power spectrum can be written as \cite{Castro05}
\begin{align}
\mathcal{C}_{\ell}^{\phi \phi} (k_1 , k_2) &= \frac{16}{\pi^2 c^4} \, \int^{\infty}_0 d k^{\prime} \, k^{\prime 2} \, \mathcal{I}_{\ell}^{\phi} (k_1 , k^{\prime}) \, \mathcal{I}_{\ell}^{\phi} (k_2 , k^{\prime}) , \\
\label{eqn:calI} \mathcal{I}^{\phi}_{\ell} (k_i , k^{\prime}) &= k_i \, \int^{\infty}_0 dr \, r^2 \, j_{\ell} (k_i r) \\ & \quad \times \int^r_0 dr^{\prime} \, F_K (r , r^{\prime} ) \, j_{\ell} (k^{\prime} r^{\prime}) \, \sqrt{P_{\Phi \Phi} (k^{\prime} ; r^{\prime}) } \nonumber .
\end{align}
\n
In both GR and modified theories of gravity we assume that the correlations of the Weyl potential $\Phi$ are significantly non-zero such that we assume the potential power spectrum is approximately \cite{Castro05}
\begin{align}
P_{\Phi_{\rm{WL}}} (k ; r , r^{\prime}) \simeq \sqrt{P_{\Phi_{\rm{WL}}} (k;r) \, P_{\Phi_{\rm{WL}}} (k ; r^{\prime})} .
\label{eqn:factorize}
\end{align}
\n
As the correlations of $\delta$ are restricted to very small scales $| \bfr - \bfr^{\prime} | \leq 100 h^{-1} \rm{Mpc}$, we simply replace $P_{\Phi_{\rm{WL}}} (k ; r , r^{\prime})$ by $P_{\Phi_{\rm{WL}}} (k ; r)$.

The lensing power spectrum can be written as
\begin{align}
P_{\Phi_{\textrm{WL}}} (k ; z) &= D^2_{+} (k ; z) \, P_{\Phi_{\textrm{WL}}}  ( k ; 0) ,
\end{align}
which, in the GR limit, can also be expressed in terms of the matter power spectrum via the Poisson equation as per Eqns. (\ref{Psi-N-def}) and (\ref{eq:GRPhi})
\begin{align}
P_{\Phi_{\textrm{WL}}} (k ; z) &\GReq \left( \frac{3 \Omega_{m0} H^2_0}{2 a k^2} \right)^2  P_{\delta \delta} (k ; z) .
\end{align}
\n
The shear lensing spectrum is related to the lensing potential spectrum via Eq. (\ref{eqn:shear_lensing_relation})
\begin{align}
\calC^{\gamma \gamma}_{\ell} (k_1 , k_2) &= \frac{1}{4} \frac{(\ell +2 )!}{(\ell - 2)!} \calC^{\phi \phi}_{\ell} (k_1 , k_2) .
\end{align}
\n
This result again assumes perfect sky coverage and neglects finite survey, selection function and sky mask effects. These may be accounted for using a pseudo-$\calC_{\ell}$ approach as detailed in \cite{Munshi10b}, where the method was used to reduce the data. We show typical sFB weak lensing spectra for $\ell = 40$ and $r_0 \in \lbrace 1400, 1800, 2200 \rbrace \, h^{-1} \rm{Mpc}$ in Figure \ref{fig:3DWLGR}. The typical deviations from GR for the f(R) models are shown in Figure \ref{fig:3DWLDifffR} and for dilaton models in Figure \ref{fig:3DWLDiffdilaton}.

\subsubsection{Noise Contributions and Systematics}
In reality, the data that we have access to are estimates of the shear field at given 3D positions in space. The radial coordinates are usually not known precisely but are estimated by some photometric redshift which implicitly has an error attached to it, typically of the order $\sigma_z \sim 0.02 - 0.1$. These photometric redshift estimates lead to a smoothing of the distribution in a radial direction. A more complete analysis would take this effect into account \cite{Heavens03}. We neglect these contributions but do include effects from shot noise under the assumption that galaxies are a Poisson sampling of an underlying smooth field \cite{Peebles80,Heavens03}. The variance of the shear estimate for a single galaxy will be dominated by the variance in intrinsic ellipticity $\sigma^2_{\epsilon}$ rather than by lensing \cite{Heavens03,Kitching08a, Kitching10}. This leads to a shot-noise term of the form 
\begin{align}
\langle \gamma_{\alpha} \gamma^{\ast}_{\beta} \rangle &= \frac{\sigma^2_{\epsilon}}{2} \delta^K_{\alpha \beta} .
\end{align}
\n
The intrinsic ellipticity signal can be difficult to separate and remains a poorly understood noise contribution. 

Fundamentally, the choice of non-linear matter power spectrum used in our analysis depends on the underlying cosmology. In this paper we aim to constrain deviations from General Relativity with respect to a fixed $\Lambda$CDM background. However, deviations induced by a modified theory of gravity are degenerate with a cosmology that incorporates baryonic feedback and/or massive neutrinos. In a 2D analysis, where we used Limber's approximation to perform a projection, we mix up the $k$ and the $\ell$ modes making any discrimination between the various non-linear effects much harder. In 3D, we can study the non-linearities mode-by-mode as a function of $k$ and $\ell$ such that it may be possible to disentangle the various non-linear contributions. The tomographic projections and their correlators can always be constructed from the 3D spherical Fourier-Bessel analysis, as the required information is already present. 

We have ignored the effect of baryonic feedback 
and massive neutrino in our study.
However, the effect of these on modification
of matter power spectrum are degenerate especially
at smaller wave numbers $k<10^{-1} h \rm{Mpc}^{-1}$
with that of MG theories.
The impact of both effects can be included as 
a multiplicative redshift and scale dependent bias. 
The feedback effect, which includes the 
effects of cooling, heating, star-formation and evolution,
chemical enrichment, supernovae feedback. 
The modelling of such processes is difficult and 
involve running hydrodynamic simulations \cite{Schaye10}.
The effect of massive neutrinos can similarly be 
incorporated using an effective multiplicative bias
(see \cite{Harnois-Deraps15} for results in projection).
Our formalism can readily be used to incorporate such effects. For example, we could schematically write \cite{Harnois-Deraps15}
\begin{align}
P_{\rm{tot}} (k,z) = P_{\rm{MG}} (k,z) b_{f}^2 (k,z) b_{\nu}^2 (k,z) ,
\end{align}
\n
and use this in the approximation given Eq.(\ref{eqn:factorize}). This would simply serve to renormalise Eq.(\ref{eqn:calI}) meaning that feedback mechanisms can be transparently folded into the analysis. A discussion of the impact of massive neutrinos on the 3D spherical Fourier-Bessel power spectrum can be found in \cite{Kitching08,DeBernardis09}.

Unless otherwise specified, the canonical survey configuration that we adopt for the 3D weak lensing spectra has a survey depth of $r_0 = 1400 h^{-1} \, \rm{Mpc}$ and takes multipoles in the range $\ell \in \lbrace 10, 80 \rbrace$ to ensure that we are in the regime that probes large scales. In this regime, we can assume contamination effects from feedback etc will be subdominant. 

\begin{figure*}[t]
\begin{center}
\includegraphics[width=150mm]{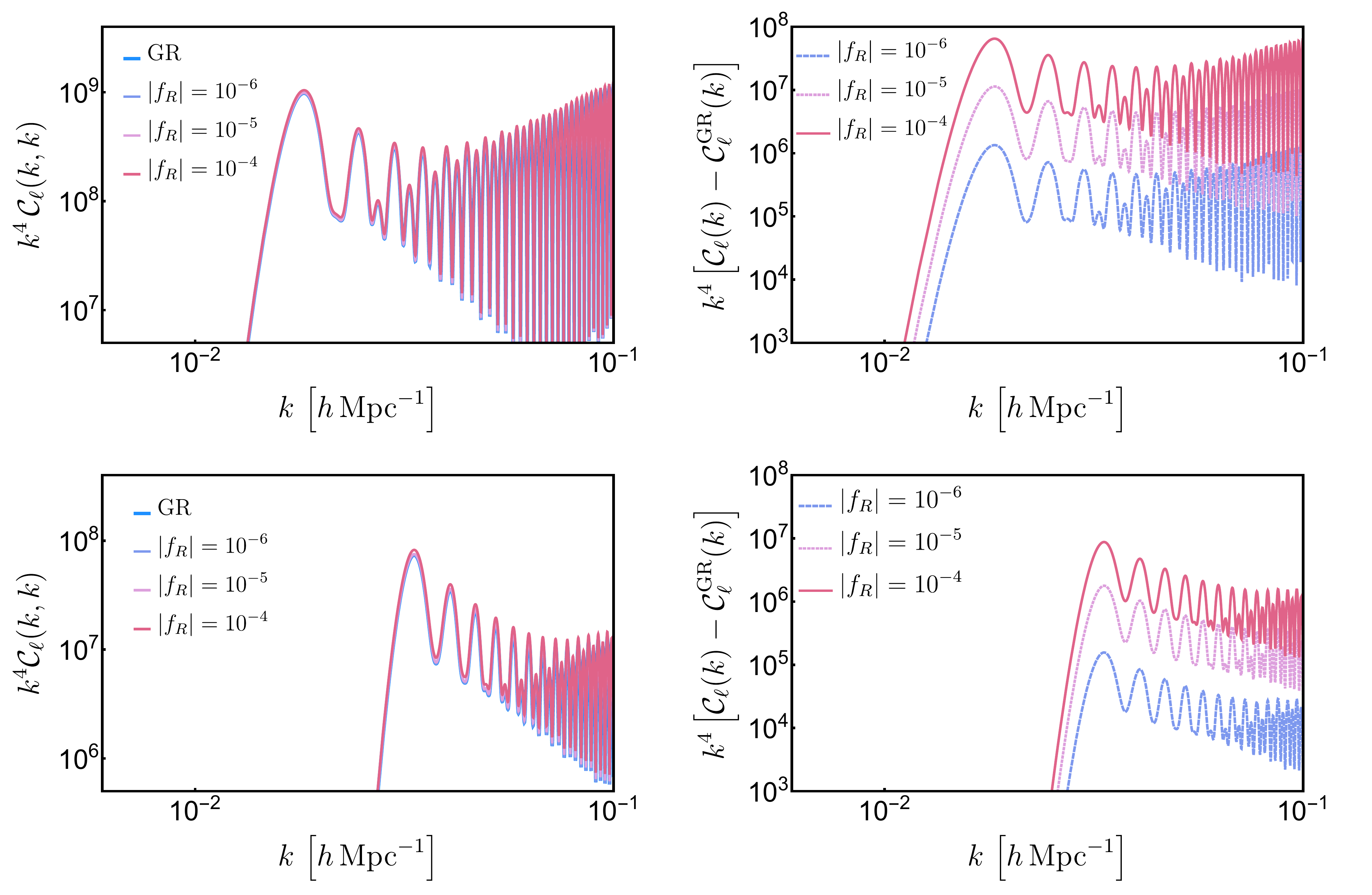}
\caption{
In this plot we show the 3D weak lensing spectra for the $f(R)$ models assuming a survey depth of $r_0 = 1400 h^{-1}$. The left plots show the total $\ell = 20$ (top) and $\ell = 40$. The plots on the right show the difference between the fiducial GR model and $|f_R | = 10^{-4}$ (solid), $10^{-5}$ (dashed) and $10^{-6}$ (dotted). All plots have been scaled by $k^4$ in order to enhance the differences. As can be seen, the signatures of a deviation from GR are most prominent in the range $k \sim 10^{-2}-10^{-1} h \rm{Mpc}^{-1}$. Scales on the order $k > 0.1 h \rm{Mpc}^{-1}$ will be probed by higher multipoles $\ell > 80$, where systematic effects and noise render our analysis and modelling insufficient and the spectra tend towards that of GR.
}
\label{fig:3DWLDifffR}
\end{center}
\end{figure*}

\begin{figure*}[t]
\begin{center}
\includegraphics[width=150mm]{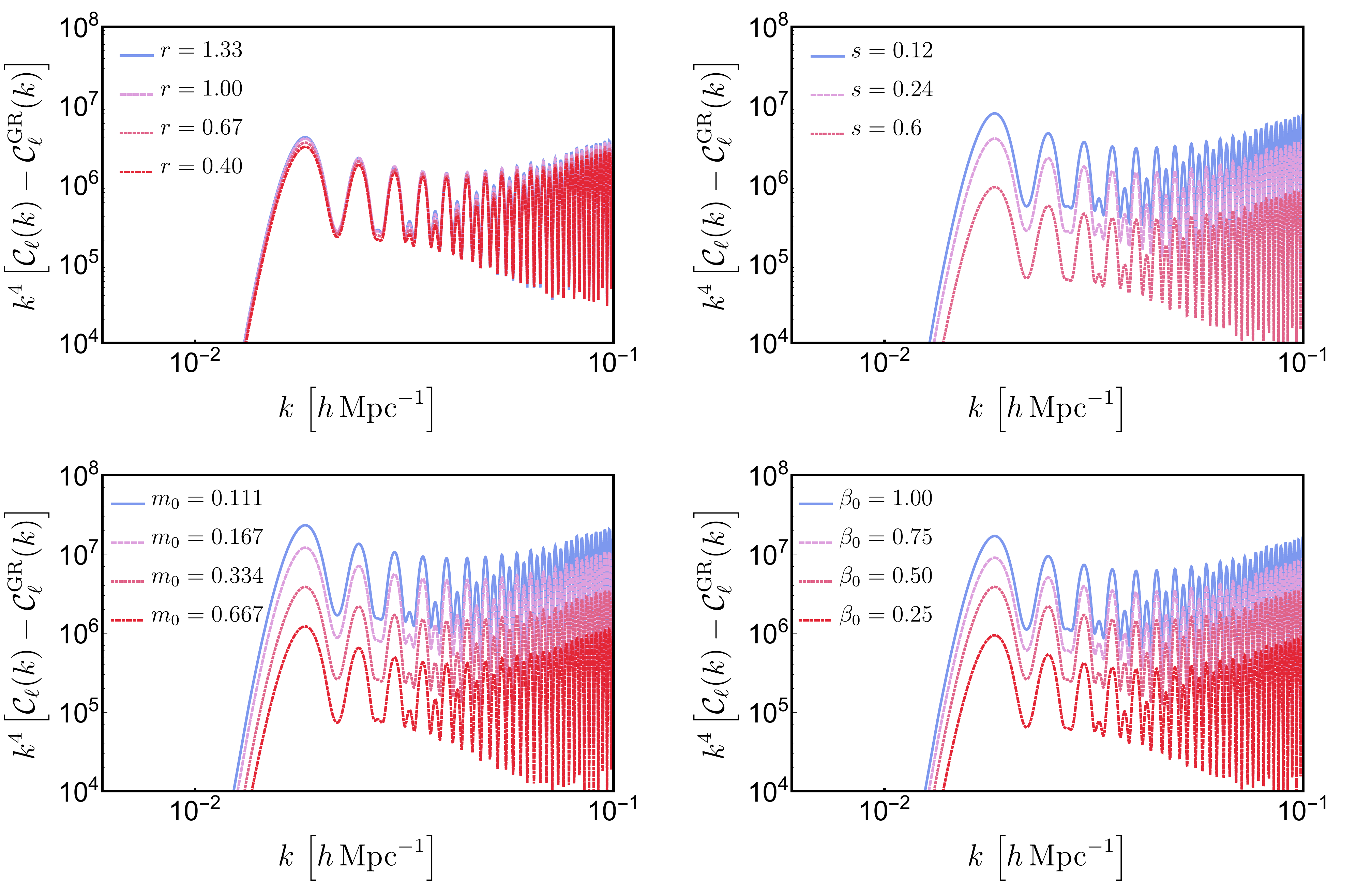}
\caption{
Here we plot the 3D weak lensing spectra for the dilaton models assuming $r_0 = 1400 h^{-1} \textrm{Mpc}$ and $\ell = 20$. These plots show the deviations from GR induced by systematically varying the model parameters $\lbrace m_0 , r , s , \beta_0 \rbrace$. The top left plot varies $r$, the top right $s$, the bottom left $m_0$ and the bottom right $\beta_0$. As before these plots are scaled by $k^4$. The fiducial model $A2$ for the dilaton spectra is characterised by $\beta_0 =  0.5$, $m_0 = 0.334$, $r = 1.00$ and $s = 0.24$. The variations in $\beta_0$ are given by the B-series, in $m_0$ by the E-series, in $r$ by the C-series and in $s$ by the A-series. As discussed previously, deviations are most significant between $k \sim 10^{-2} - 10^{-1} h \rm{Mpc}^{-1}$ probed by multipoles $\ell < 100$. 
}
\label{fig:3DWLDiffdilaton}
\end{center}
\end{figure*}

\section{Statistical Analysis}
\label{sec:5}
\subsection{Comparison with Previous Studies}
In this section we briefly discuss the plethora of constraints on modified theories of gravity and how competitive we expect 3D weak lensing to be. The constraints use multiple probes that fold in information from a wide range of astrophysical processes covering a very broad range of scales. Current tests of GR on cosmological scales are still not at the levels of precision offered by small scale experiments. Some of these constraints, such as weak lensing, redshift space distortions and galaxy clustering, require us to fix a particular model in order to estimate and reconstruct the growth rate of large scale structure from data. In addition, many of the observable effects are vexed by a poor understanding of baryonic physics, such as the role of non-linear galaxy bias or feedback mechanisms in both standard GR and modified theories of gravity, necessarily introducing uncertainty into the constraints. Here we show how 3D weak lensing at low multipoles, $\ell < 100$, is likely to be a relatively clean probe of modified theories of gravity with systematic effects and constraints only becoming significant at higher multipoles.  

\subsubsection{$f(R)$ Models}
Amongst the largest scales we can probe, CMB lensing from Planck offers the deepest line of sight constraints on modified theories of gravity but only places a relatively weak constraint of $|f_R| \leq 10^{-2}$ for $z \lesssim 6$ \cite{PlanckMG}. On similarly large scales, it is highly anticipated that a joint analysis of 21cm intensity mapping and the CMB could yield much tighter constraints of $|f_R| \leq 10^{-5}$ for $z \sim 0.7-2.5$ due to the greater number of useful Fourier modes \cite{Hall13}. Constraints from the galaxy power spectrum as measured by WiggleZ yielded a relatively good constraint on the order of $|f_R| \leq 1.4 \times 10^{-5}$ $95$\% at $z \sim 0.2-1$ \cite{Dossett14}. A recent galaxy clustering ratio $\eta$ was proposed as a means to avoid a number of systematic uncertainties associated to scale-dependent growth rates. This ratio applied to SDSS data yielded a constraint of $|f_R| < 4.6 \times 10^{-5}$ at the $95$\% confidence limit \cite{Bel15}. Redshift space distortions measured in the range $z \sim 0.16 - 0.47$ yield $|f_R| \leq 10^{-4}$ \cite{Yamamoto10}. Recent CFHTSLenS 2D weak lensing results lead to relatively weak constraints of $|f_R| \leq 10^{-4}$ \cite{Harnois-Deraps15}.  

On smaller scales, Coma gas measurements yield $|f_R| \leq 6 \times 10^{-5}$ for $z \sim 0.02$ \cite{Terukina13}, strong lensing of galaxies on kpc scales places a constraint of $|f_R| \leq 2.5 \times 10^{-6}$ \cite{Smith09} and solar system scales typically introduce constraints on the order of $|f_R| \leq 8 \times 10^{-7}$ \cite{Hu07}.

\subsubsection{Dilaton Models}

CFHTSLenS weak lensing places rather broad constraints on the dilaton models with the data preferring lower values of $s, r$ and $\beta_0$ \cite{Harnois-Deraps15}. The analysis of the diagonal direction in the $\lbrace m_0 , s \rbrace$ plane prefers lower values of $m_0$. A recent study by \cite{Hojjati15} using LSST, weak lensing, galaxy clustering and Planck CMB data was able to place $1 \sigma$ constraints on $\xi_0 = H_0 / ( c m_0)$ and $\beta_0 = 1$ of $2.7 \times 10^{-5}$ and $2.3 \times 10^{-1}$ respectively. When $\beta_0 = 5$, it was found that $\xi_0 < 3 \times 10^{-3}$ at $95$\% CL. It was also shown that if $\beta_0 \sim 1$ then current data cannot place any meaningful constraints on $\xi_0$, however when folding in LSST data the constraint tightens to $\xi_0 \sim \rm{few} \times 10^{-5}$ and it should be possible to measure $\beta_0 \sim 1$ to within $20$\% accuracy. 

\subsection{$\chi^2$ Analysis}
\subsubsection{Overview}
Following \cite{Munshi15}, we can define a likelihood $\mathcal{L}$ for an arbitrary set of parameters $\theta_{\alpha}$ specifying our modified theory of gravity, i.e. we implicitly assume a fixed background cosmology. For $f(R)$ theories the parameter vector is just $\theta_{\alpha} = \lbrace n , f_{R_0} \rbrace$. For the dilaton theories, the parameter vector is given by $\theta_{\alpha} = \lbrace s , \beta_0 , r , m_0 \rbrace$. Given a noisy data vector $\tilde{\mathcal{C}}_{\ell} (k)$ the likelihood is given by
\begin{widetext}
\begin{align}
\mathcal{L} (\theta_{\alpha} | \tilde{\calC}_{\ell} (k) ) &= \frac{1}{( 2 \pi )^{N_{\textrm{pix}} / 2} \, | \textrm{det} \, \mathbb{C} |^{1/2}} \, \exp \, \left[ - \frac{1}{2} \displaystyle\sum_{\ell \ell^{\prime}} \int dk \int dk^{\prime} \, \delta \mathcal{C}_{\ell} (k) \, \mathbb{C}^{-1}_{\ell \ell^{\prime}} (k,k^{\prime}) \, \delta \mathcal{C}_{\ell^{\prime}} (k^{\prime}) \right] .
\end{align}
\end{widetext}
\n
Here $\delta \tilde{\mathcal{C}}_{\ell} = \tilde{\mathcal{C}}_{\ell} - \tilde{\mathcal{C}}^{\textrm{GR}}_{\ell}$, $N_{\textrm{pix}}$ is the size of the data vector and $\mathbb{C}_{\ell \ell^{\prime}} (k, k^{\prime})$ is the covariance matrix defined by
\begin{align}
\mathbb{C}_{\ell \ell^{\prime}} (k,k^{\prime}) &= \frac{2}{2 \ell + 1} \, \left[ \tilde{\calC}_{\ell} (k,k^{\prime}) +N_{\ell} (k,k^{\prime}) \right]^2 \, \delta^K_{\ell \ell^{\prime}} \, \delta^K_{m m^{\prime}} ,
\end{align}
\n
where we have made use of the Gaussian approximation and $N_{\ell} (k,k^{\prime})$ is a noise term. If we have all-sky coverage then the covariance matrix reduces to a block-diagonal form with the matrix being diagonally dominated in the $\lbrace k , k^{\prime} \rbrace$ space. Introducing a sky-mask or assuming partial sky-coverage will induce mode-coupling terms between the harmonics that results in off-diagonal terms. The $\chi^2$ statistic is given by
\begin{align}
\chi^2 &= \displaystyle\sum_{\ell \ell^{\prime}} \int dk \, \int dk^{\prime} \, \delta \mathcal{C}_{\ell} (k) \, \mathbb{C}^{-1}_{\ell \ell^{\prime}} (k,k^{\prime}) \, \delta \mathcal{C}_{\ell^{\prime}} (k^{\prime}) .
\end{align}
\n
In order to constrain the parameters of the modified theory of gravity, we assume a perfect knowledge of the background LCDM cosmology. We assume multipoles in the range $\ell \in \lbrace 20, 80 \rbrace$ which corresponds to probing the scales corresponding to $k \in \lbrace 10^{-2}, 10^{-1} \rbrace \, h \, \rm{Mpc}^{-1}$. In Figure \ref{fig:chi2fR} we demonstrate how the $\chi^2$ constraints vary with multipoles $\ell$ used, survey depth $r_0$ and ellipticity variance $\sigma_{\epsilon}$. By folding in more multipoles we can naturally gain tighter constraints. Increasing the survey depth allows us to probe a greater range of redshifts, also tightening the constraints. Similarly, increasing the ellipticity variance naturally degrades our constraints due to the increased noise. We only consider the multipoles $\ell \leq 80$ as high multipoles probe scales $k > 0.1 h \rm{Mpc}^{-1}$, where the spectra tend to GR quickly and systematic effects can become substantial. The assumed galaxy number density is fixed at $\bar{N} = 10^{-4} \textrm{Mpc}^{-3}$ in order to compare our results to those of \cite{Munshi15}. 

\subsubsection{Constraints From $\chi^2$}
Our constraints are derived assuming a prototypical 3D weak lensing survey of $r_0 = 1400  h^{-1} \, \rm{Mpc}$, taking multipoles $\ell \in \lbrace 10,80 \rbrace$ and taking wavenumbers in the range $k \in \lbrace 10^{-2}, 10^{-1} \rbrace \, h \, \rm{Mpc}^{-1}$. We assume an ellipticity variance of $\sigma_{\epsilon} = 0.2$. 

For the $f(R)$ models, the $3\sigma$ limits are $| f_R | \lesssim 5 \times 10^{-6}$ for the $n = 1$ models and $| f_R | \lesssim 9 \times 10^{-6}$ for the $n = 2$ models. The results are shown in Figure \ref{fig:chi2fR}. 

The total 3$\sigma$ constraints for the dilaton model with parameters $\lbrace m_0 , r , \beta_0 , s \rbrace$ about a fiducial model of $\lbrace 0.334, 1.0, 0.5, 0.24 \rbrace$ are $\lbrace 0.47, 0.85, 0.38, 0.39  \rbrace$ as shown in Figure \ref{fig:chi2dil}. This means that should we observe a 3D weak lensing spectra consistent with GR, the fiducial model $A_2$ can be excluded at the $3\sigma$ level and the data should be capable of ruling out variations of the fiducial model with values of $m_0 < 0.47$, $\beta_0 > 0.38$ and $s < 0.39$. The analysis seems to be indiscriminate towards $r$ with the $\chi^2$ values being relatively flat across the parameter space. 

\begin{figure*}[t]
\begin{center}
\includegraphics[width=150mm]{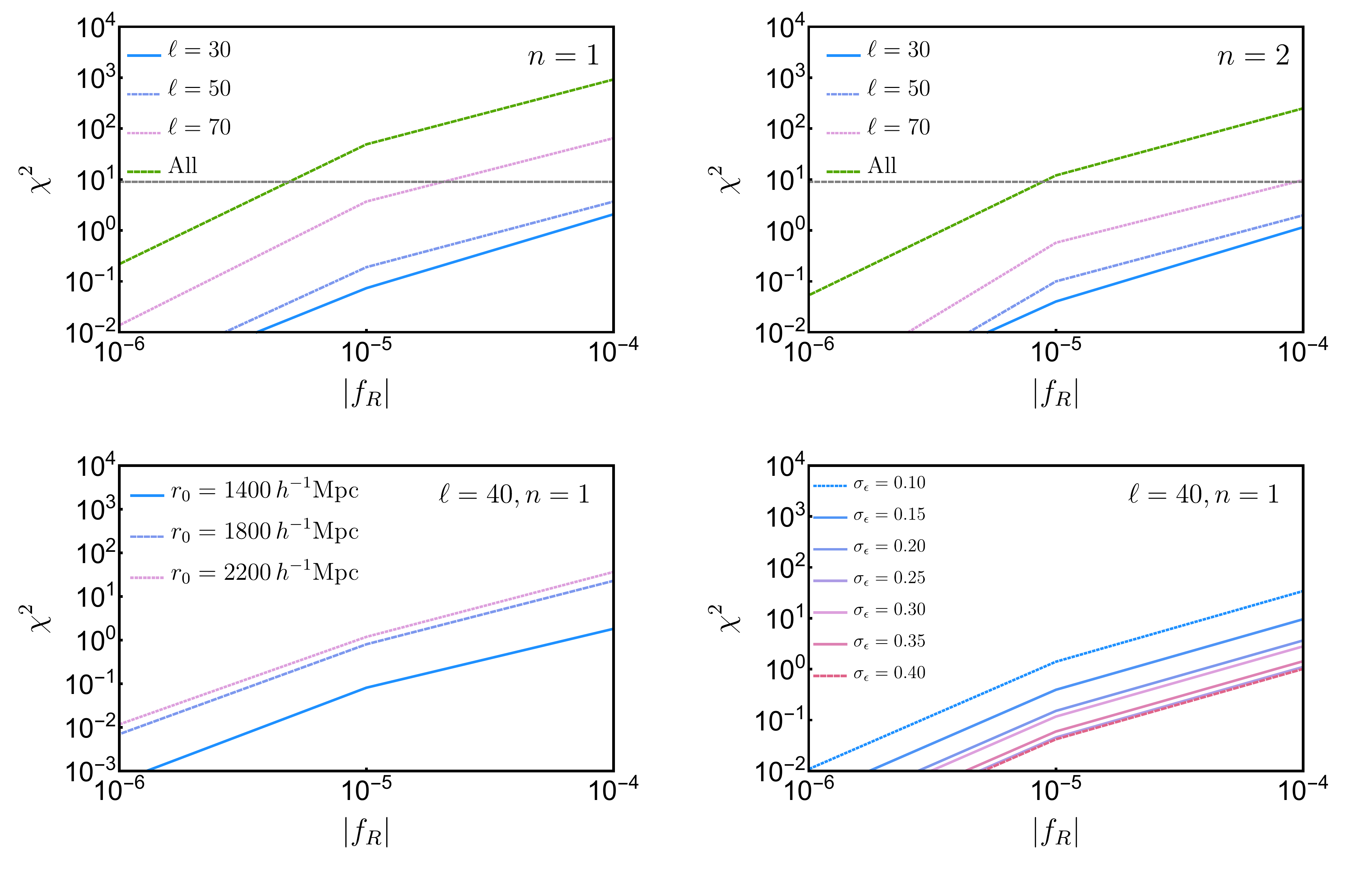}
\caption{
Here we show a $\chi^2$ analysis for the $f(R)$ models as a function of multipole $\ell$, survey depth $r_0$ and ellipticity variance $\sigma_{\epsilon}$. We use all multipoles up to $\ell \sim 70$, take survey sizes $r_0 \in \lbrace 1400, 1800, 2200 \rbrace \, h^{-1} \textrm{Mpc}$ and assume an ellipticity variances in the range $\sigma_{\epsilon} = \lbrace 0.1 , 0.4 \rbrace$. In the top two plots we set $r_0 = 1400 h^{-1} \textrm{Mpc}$ and $\sigma_{\epsilon} = 0.2$ and demonstrate how individual multipoles ($30,50,70$) contribute to $\chi^2$ compared to the sum of all the multipoles in the range $\ell \in \lbrace 30,70 \rbrace$ (top, green curve). The top left plot is for $n=1$ and the top right for $n=2$. In the bottom left plot we adopt fiducial values of $\ell = 40$, $\sigma_{\epsilon} = 0.2$ and vary the survey depth $r_0 \in \lbrace 1400, 1800, 2200 \rbrace h^{-1} \rm{Mpc}$ (top curve to bottom curve). For bottom right plot, we adopt a fiducial survey of $\ell = 40$ and $r_0 = 1400 h^{-1} \textrm{Mpc}$ for varying $\sigma_{\epsilon}$. As anticipated, stronger noise degrades the constraints. The data is binned in the range $k \sim 0.01 - 0.1 \, h \, \textrm{Mpc}^{-1}$. At the $3\sigma$ level we find constraints on the $n=1$ models to be $\sim | f_R | < 5 \times 10^{-6}$ and for the $n = 2$ models this is degraded to $\sim |f_R| < 9 \times 10^{-6}$.
}
\label{fig:chi2fR}
\end{center}
\end{figure*}

\begin{figure*}[t]
\begin{center}
\includegraphics[width=150mm]{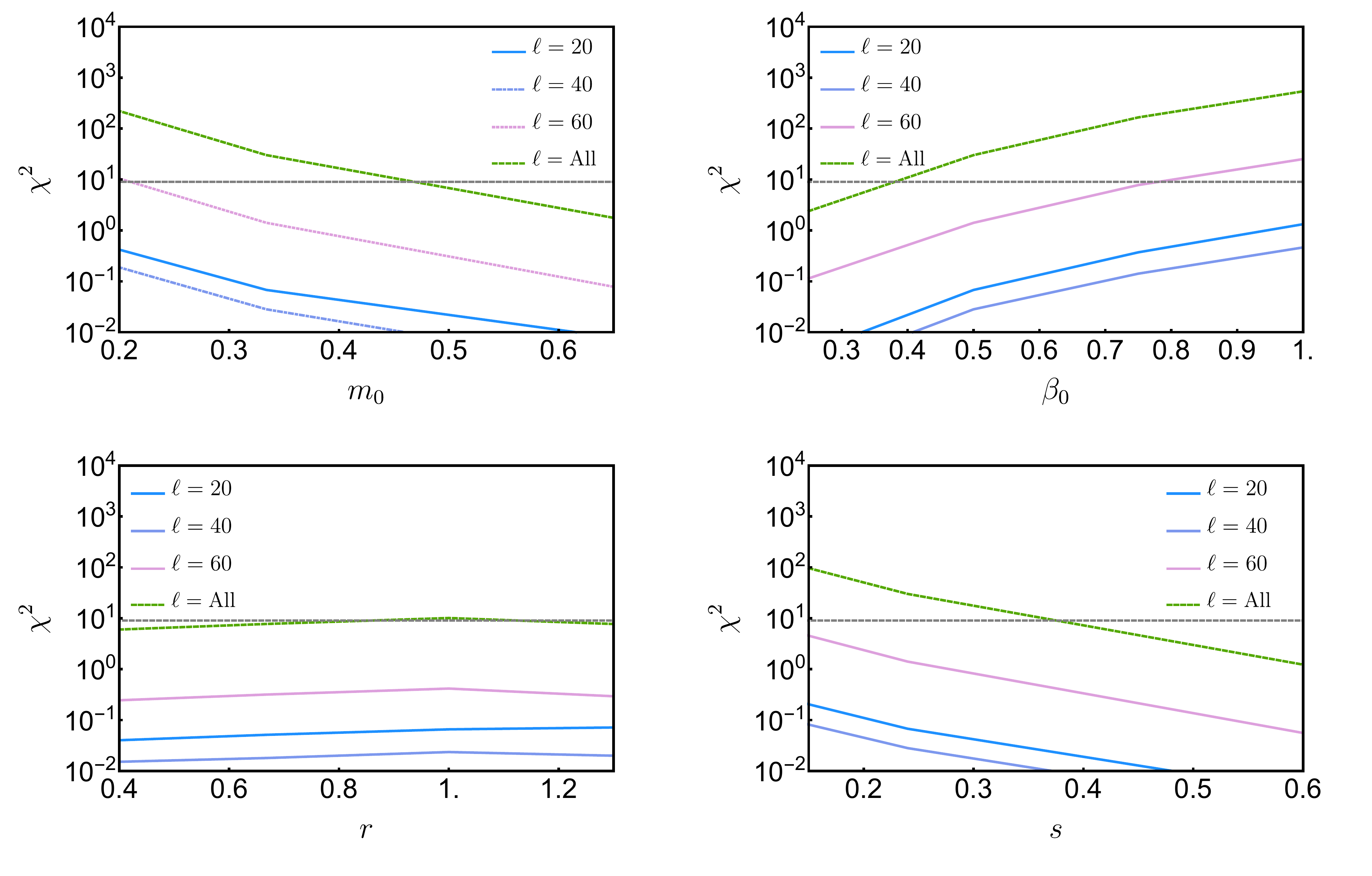}
\caption{
A $\chi^2$ analysis for the dilaton models parameterised by $\lbrace m_0 , \beta_0 , r , s \rbrace$. Here we take multipoles up to $\ell = 60$, a survey depth of $r_0 = 1400 h^{-1} \textrm{Mpc}$ and assume an ellipticity variance of $\sigma_{\epsilon} = 0.2$. The $3\sigma$ constraints on the various parameters are: $m_0 \sim 0.4$, $\beta_0 \sim 0.425$ and $s \sim 0.37$. For $r$ we cannot place a useful constraint on the parameter. For ellipticity variances on the order $\sim 0.3$ and including photometric redshift smoothing, it is very likely that we would lose any statistical significance on $r$. 
}
\label{fig:chi2dil}
\end{center}
\end{figure*}

\subsection{Fisher Matrix Analysis}
\subsubsection{3D Fisher Matrix}
The Fisher information matrix is a useful, though limited, tool in modern cosmology. Assuming that the likelihood surface near the peak is sufficiently well approximated by a multivariate Gaussian, the Fisher matrix (FM) will inform us about the Gaussian uncertainties at a given point in parameter space. At a basic level, it allows us to place a theoretical lower limit on the parameter space uncertainties. There is, however, no good way to include prior information and nor is there a means to perform a global exploration of a parameter space that may contain multiple peaks or other features, such as walls. Consequentially, the Fisher information matrix can occasionally yield some rather suspect cross-correlations. Some of these issues may be solved by resorting to a full MCMC sampling of the parameter space, though we leave such an implementation to the future. In this paper, we focus on placing a naive theoretical $1\sigma$ bound on the model parameters for the modified theories of gravity considered here. In particular, we will focus on how the $1\sigma$ bound varies as a function of the survey configuration, multipoles used and ellipticity variance.  
A useful discussion of Fisher matrix forecasting in the sFB formalism may be found in \cite{Lanusse14,Nicola14}.  Our discussion will closely follow the methods outlined in these papers. 
The Fisher matrix (FM) is defined to be the inverse covariance matrix of the posterior distribution
\begin{align}
F_{\alpha \beta} &= \left\langle \frac{\partial^2 \mathcal{L}}{\partial \theta_{\alpha} \partial \theta_{\beta} } \right\rangle ,
\end{align}
\n
where $\mathcal{L} = - \ln L$ is the log-likelihood. If all parameters bar one are fixed, then the \textit{fixed uncertainty} will be bounded by
\begin{align}
\Delta \theta_{\alpha} &\geq \frac{1}{\sqrt{F_{\alpha \alpha}}} .
\end{align}
\n
However, if several parameters are estimated from our data then the minimum standard deviation is given by
\begin{align}
\Delta \theta_{\alpha} &\geq \left( F^{-1} \right)^{1/2}_{\alpha \alpha} ,
\end{align}
\n
this is called the \textit{marginalised uncertainty}. The Fisher matrix for the sFB spectra may be computed via a number of different implementations. The calculation of the FM typically begins by assuming a Gaussian likelihood for the sFB harmonics but quickly becomes complicated due to the complex correlations between the various $k$-modes induced by both finite survey and masking effects as well as by the intrinsic time-evolution of the underlying random field. In our case, this will be the time evolution of the lensing potential. 

The non-diagonal correlations can be dealt with via two different approaches: 1) we choose a finite grid in $k$ space and compute the FM on the discrete grid or 2) we approximate the full FM by a diagonal data covariance matrix evaluated at discrete points $k_i$ \cite{Nicola14}. If the covariance matrix can be well approximated as diagonally dominant, then care must be taken in the choice of bin size as bins that are too small would overestimate the information content as we neglect correlations between neighbouring wavenumbers. For a discrete grid, smaller bin sizes $\Delta k_i$ mean that the covariance matrix becomes increasingly complex to invert whereas larger bin sizes would be tantamount to discarding information \cite{Lanusse14}. As discussed in \cite{Lanusse14}, care must also be taken when choosing the largest scale $k_{\textrm{min}}$ to include in the analysis, as at small $k$ the sFB spectra can become extremely small but it is possible that they can still contribute to the Fisher information. 

Another problem that arises when calculating the Fisher matrices is that the condition number\footnote{$\kappa ( {\bf{A}} ) = \left| \lambda_{\rm{max}} ({\bf{A}}) / \lambda_{\rm{min}} ({\bf{A}}) \right|$.} of the covariance matrices can be quite high, making the matrices ill-conditioned. This means that if the bin size is taken to be too small then the matrix can become singular. Care must also be taken in binning the data in order to avoid numerical instabilities in the highly oscillatory regime at large $k$.

Assuming a likelihood with covariance matrix $\bfC$ and mean $\bfmu$, the FM will be given by \cite{Tegmark96,Tegmark97,Heavens03,Lanusse14,Nicola14}
\begin{align}
F_{\alpha \beta} &= \frac{f_{\textrm{sky}}}{2} \textrm{Tr} \left[ \bfC^{-1} \, \frac{\partial \bfC}{\partial \theta_{\alpha}} \, \bfC^{-1} \, \frac{\partial \bfC}{\partial \theta_{\beta}} \right] + \frac{\partial \bfmu^T}{\partial \theta_{\alpha}} \, \bfC^{-1} \, \frac{\bfmu}{\partial \theta_{\beta}} .
\label{eqn:Fisher}
\end{align}
\n
If there is no angular mask, the sFB coefficients will remain uncorrelated between different multipoles $\ell$ and the covariance matrix $\bfC$ reduces to a block diagonal form. A further simplification arises as $\mu = \langle \zeta_{\ell m} (k) \rangle = 0$, where $\zeta$ is defined in Eq. (\ref{eqn:zeta}), and the FM reduces even further to the following form
\begin{align}
F_{\alpha \beta} &= f_{\textrm{sky}} \displaystyle\sum_{\ell} \, \frac{(2 \ell + 1) \, \Delta \ell}{2} \, \textrm{Tr} \, \left[ \bfC^{-1}_{\ell} \, \frac{\partial \bfC_{\ell}}{\partial \theta_{\alpha}} \, \bfC^{-1}_{\ell} \, \frac{\partial \bfC_{\ell}}{\partial \theta_{\beta}} \right] .
\label{eq:FMFull}
\end{align}
Schematically, the data covariance matrix for the sFB spectra measured at a set of discrete radial wavenumbers $k_i$ can be written as \cite{Nicola14,Lanusse14}
\begin{align}
\bfC_{\ell} = \begin{pmatrix}
\tilde{\calC}_{\ell} (k_0,k_0) & \tilde{\calC}_{\ell} (k_0,k_1) & \dots & \tilde{\calC}_{\ell} (k_0,k_{\textrm{max}}) \\
\tilde{\calC}_{\ell} (k_1,k_0) & \tilde{\calC}_{\ell} (k_1,k_1) & \dots & \tilde{\calC}_{\ell} (k_1,k_{\textrm{max}}) \\
\vdots & \vdots & \ddots & \vdots \\
\tilde{\calC}_{\ell} (k_{\textrm{max}},k_0) & \tilde{\calC}_{\ell} (k_{\textrm{max}},k_1) & \dots & \tilde{\calC}_{\ell} (k_{\textrm{max}},k_{\textrm{max}})
\end{pmatrix} ,
\end{align}
\n
where, as before,
\begin{align}
\tilde{\calC}_{\ell} (k_i,k_j) &= \calC_{\ell} (k_i , k_j ) + N_{\ell} (k_i , k_j ).
\end{align}

\subsubsection{Constraints from Fisher Matrix}
For a survey of depth $r_0 = 1400 h^{-1} \rm{Mpc}$, ellipticity variance of $\sigma_{\epsilon}$ = 0.2 and at a fixed multipole of $\ell = 20$, the $1\sigma$ fractional error on $|f_R|$ is typically $\Delta | f_R | / f_R \sim 0.0214 \, (0.0240)$ for $n=1 \, (n=2)$. This means that an optimistic weak lensing survey should be able to measure $|f_R|$ to the percent level accuracy, with accuracies below $10$\% at the 3$\sigma$ confidence limit. 

The $1\sigma$ fractional errors $\Delta \theta_{\alpha} / \theta_{\alpha}$ on the dilaton parameters $\lbrace m_0 , \beta_0 , r , s \rbrace$ are on the order of $\sim$ $\lbrace 0.0383, 0.0790, 0.4480, 0.3267 \rbrace$ for $\ell = 20$ and $\sigma_{\epsilon} = 0.2$. The dependence of the $1\sigma$ errors estimated from the Fisher matrix on the ellipticity variance is shown in Figure \ref{fig:dil_onesigma_ellipticity} for the dilaton models. 

As discussed in \cite{Heavens03}, the 3D modes themselves are generally noisy but by improving the survey characteristics we can include more effective 3D modes with good signal-to-noise ratios. Data compressions techniques, such as the Karhunen-Lo\`eve analysis used in \cite{Heavens03} would be a step towards reducing the size of the data sets whilst having as little impact on estimated errors of the model parameters.  

These constraints derived here assume a sky-fraction of $f_{\rm{sky}} = 1$, with the errors scaling as $f^{-1/2}_{\rm{sky}}$. Improvements to these constraints will typically depend on the characteristics of the survey. For instance, reducing the ellipticity variance, increasing the number density of source galaxies or a large sky fraction will all improve the constraints. Likewise, may also fold in higher multipoles in order to improve the constraints. We leave a detailed study of optimal survey configurations to a future paper. 

It should be noted that the results presented here are optimistic as we have neglected noise contributions, such as photometric redshift errors, sky masking and sky fractions. However, the general results suggest that 3D weak lensing should be a powerful tool for upcoming large scale structure surveys, in agreement with recent studies in the literature \cite{Heavens03,Heavens07,Kitching08,Kitching08a,Kitching10,Munshi10a,Munshi10b,Camera11,Kitching14,Kitching15}.

\begin{figure}[t!]
\begin{center}
\includegraphics[width=80mm]{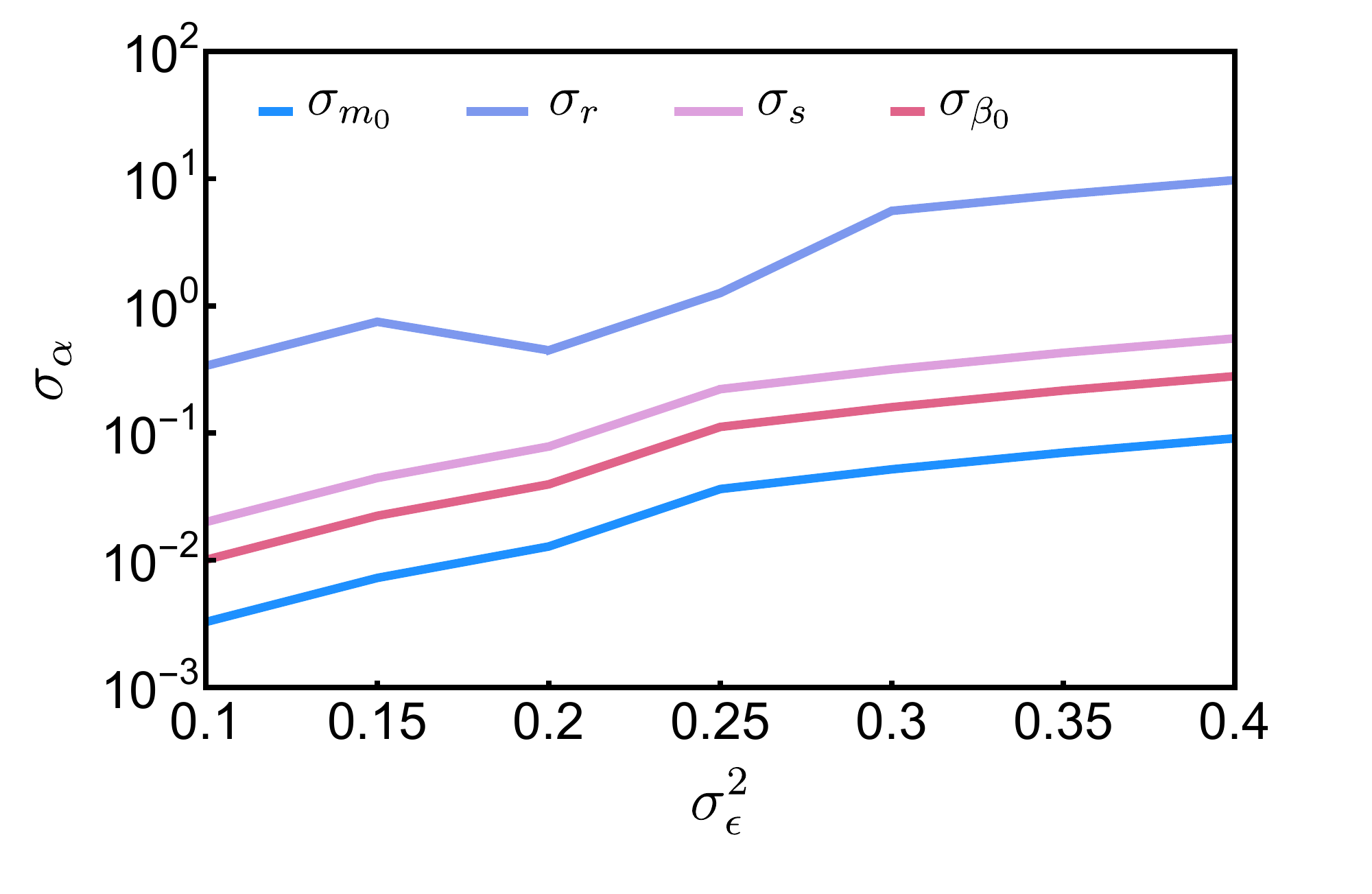}
\caption{
An example of the $1\sigma$ fixed uncertainty on the dilaton parameters as estimated from the full non-diagonal Fisher matrix $F_{\alpha \beta}$, $\Delta \theta_{\alpha} \geq ( F_{\alpha \alpha} )^{-1/2}$ for a single multipole $\ell = 20$ with a survey depth of $r_0 = 1400 h^{-1} \rm{Mpc}$. We vary the ellipticity noise $\sigma^2_{\epsilon}$. As expected, decreasing the noise leads to a decrease in the uncertainty in the terms we are sensitive to, i.e. $m_0$ and $\beta_0$, and a smaller decrease in uncertainty on the terms to which we have less sensitivity, i.e. $r$ and $\beta_0$. The lines correspond to $r$ (top), $s$, $\beta_0$ and $m_0$ (bottom) respectively. 
}
\label{fig:dil_onesigma_ellipticity}
\end{center}
\end{figure}

{\textcolor{black}{\subsection{Principal Component Analysis}}}
In this section we implement a principal component analysis (PCA) of the Fisher matrices in order to assess the accuracy to which linear combinations of model parameters may be determined from a prototypical 3D weak lensing survey. A PCA is an efficient method for determining the degeneracy directions and linear combinations of parameters, ranking them according to how accurately we may determine them from the data. This method has been applied to various cosmological data sets \cite{Efstathiou98,Efstathiou01,Munshi05} and this section will closely follow the treatment presented in \cite{Munshi05}. 

The inverse of the Fisher matrix yields the covariance of the parameter vector at the maximum likelihood
\begin{align}
\bfF^{-1} &= \langle \Delta \boldsymbol{\theta} \Delta \boldsymbol{\theta}^{T} \rangle = \langle \boldsymbol{\theta} \boldsymbol{\theta}^T \rangle - \langle \boldsymbol{\theta} \rangle \langle \boldsymbol{\theta}^T \rangle .
\end{align}
\n
The standard deviation of the $i$-th parameter is obtained from the inverse Fisher matrix via $\Delta \theta_{\alpha} = \left[ \left( \bfF^{-1} \right)_{\alpha \alpha} \right]^{1/2}$, this is the minimum variance bound (MVB). According to the Cram\'er-Rao inequality, the variance of any unbiased estimator is always larger or equal to the MVB.

Any real matrix $\bfW$ is known as a decorrelation matrix if it obeys
\begin{align}
\bfF = \bfW^T \, \boldsymbol{\Lambda} \, \bfW ,
\label{eq:EVD}
\end{align}
\n
where $\boldsymbol{\Lambda}$ is the diagonal matrix. The quantities $\boldsymbol{\Phi} = \bfW \, \boldsymbol{\theta}$ are said to be decorrelated as their covariance matrix will be  diagonalised \cite{Hamilton00}
\begin{align}
\langle \Delta \boldsymbol{\Phi} \, \Delta \boldsymbol{\Phi}^T \rangle &= \bfW \, \langle \Delta \boldsymbol{\theta} \, \Delta \boldsymbol{\theta}^T \rangle \, \bfW^T = \boldsymbol{\Lambda}^{-1}.
\end{align}
\n
We can always scale the quantities $\boldsymbol{\Phi}$ to unity variance by introducing $\tilde{\bfW} = \boldsymbol{\Lambda}^{1/2} \, \bfW$ without loss of generality . The Fisher matrix in terms of $\tilde{\bfW}$ reduces to \cite{Munshi05}
\begin{align}
\bfF &= \tilde{\bfW}^T \, \tilde{\bfW}. 
\label{eq:FWtW}
\end{align} 
\n
However, the choice of $\tilde{\bfW}$ is not unique. In particular, for any matrix $\tilde{\bfW}$ that satisfies Eq. (\ref{eq:FWtW}), the same will be true for any orthogonal rotation ${\bf{O}} \bfW$ such that ${\bf{O}} \in {\rm{SO}}(n)$ \cite{Hamilton00}. This implicitly means that there are infinitely many decorrelation matrices that satisfy Eq. (\ref{eq:FWtW}). 

Given that $\bfW$ is an orthogonal matrix, its rows will be eigenvectors $p_{\alpha}$ of $\bfF$ and the diagonal matrix of the corresponding eigenvalues will be given by $\boldsymbol{\Lambda} = \rm{diag} (\lambda_{\alpha})$. In this case, the decomposition given by Eq. (\ref{eq:EVD}) is known as a principal component decomposition (PCD) \cite{Munshi05}. The principal components of the system are given by
\begin{align}
\mu_{\alpha} = \displaystyle\sum_{\beta} \Lambda_{\alpha \beta} \, \theta_{\beta}. 
\end{align}

In a PCD, the eigenvectors of $\bfF$ will determine the principal axes of the n-dimensional error ellipsoid in the parameter manifold. The eigenvectors correspond to orthogonal linear combinations of the physical parameters that may be determined independently from the data. If these vectors are aligned with the parameter axes, then they are said to be less degenerate between those parameters. The accuracy to which the linear combination of these parameters may be determined can be quantified by the variance $\sigma_{\alpha} = \sigma (p_{\alpha}) = \lambda^{-1/2}_{\alpha}$. A given principal component of the Fisher matrix tells us how accurately a specific linear combination of parameters may be determined from the data. By convention, we will take the eigenvalues to be in descending order such that the first eigenvector $p_1$ will have the smallest variance and will therefore be the best constrained parameter combination. Similarly, the last eigenvector $p_n$ will be the direction with the greatest uncertainty. A caveat to the analysis in this section is that we implicitly assume a fixed $\Lambda$CDM background and vary the modified theory of gravity parameters about this background. In reality, it is likely that a number of the parameters introduced by a modification to gravity will be degenerate with the $\Lambda$CDM parameters. We leave a more complete analysis of this point to future work. 

The MVB may be estimated from the eigenvectors and eigenvalue of the Fisher matrix via \cite{Munshi05}
\begin{align}
\Delta \theta_{\alpha} &= \left( \displaystyle\sum^n_{i=1} W^2_{\alpha \beta} \, \lambda^{-1}_{\alpha} \right)^{1/2} .
\label{eqn:MVB}
\end{align}

For the dilaton models, we will consider the Fisher matrix $\bfF$ corresponding to all four model parameters $\lbrace m_0 , \beta_0 , r , s \rbrace$. We will study how the variance of the linear combination of parameters is sensitive to ellipticity dispersion $\sigma_{\epsilon}$, the survey depth $r_0$ and the multipoles $\ell$ used in the analysis. As would be anticipated, the noise variance induces a scaling of the parameter variance. If we neglect boundary and discretisation effects, then the covariance will be anti-proportional to the sky fraction. Although we typically set $f_{\rm{sky}}$ to unity for convenience, we note that the variance $\sigma$ scales as approximately $f^{-1/2}_{\rm{sky}}$. 

In Figure \ref{fig:PCA_dilaton}, we show the variance $\sigma (p_{\alpha})$ associated to the principal components of the Fisher matrix for the dilaton models. The first eigenvector can be seen to be dominant with the first principal component reading  
\begin{align}
\mu_1 \propto -0.9394 \, m_0 + 0.02694 \, r + 0.3051 \, \beta_0 - 0.1538 \, s,
\end{align}
\n
as per Appendix \ref{App:PCA}. The PCA essentially finds that the best measured quantities are dominated by the mass $m_0$ and the coupling $\beta_0$, in agreement with the $\chi^2$ analysis. 
\n
\phantom{TEXT}

\begin{figure}[t!]
\begin{center}
\includegraphics[width=80mm]{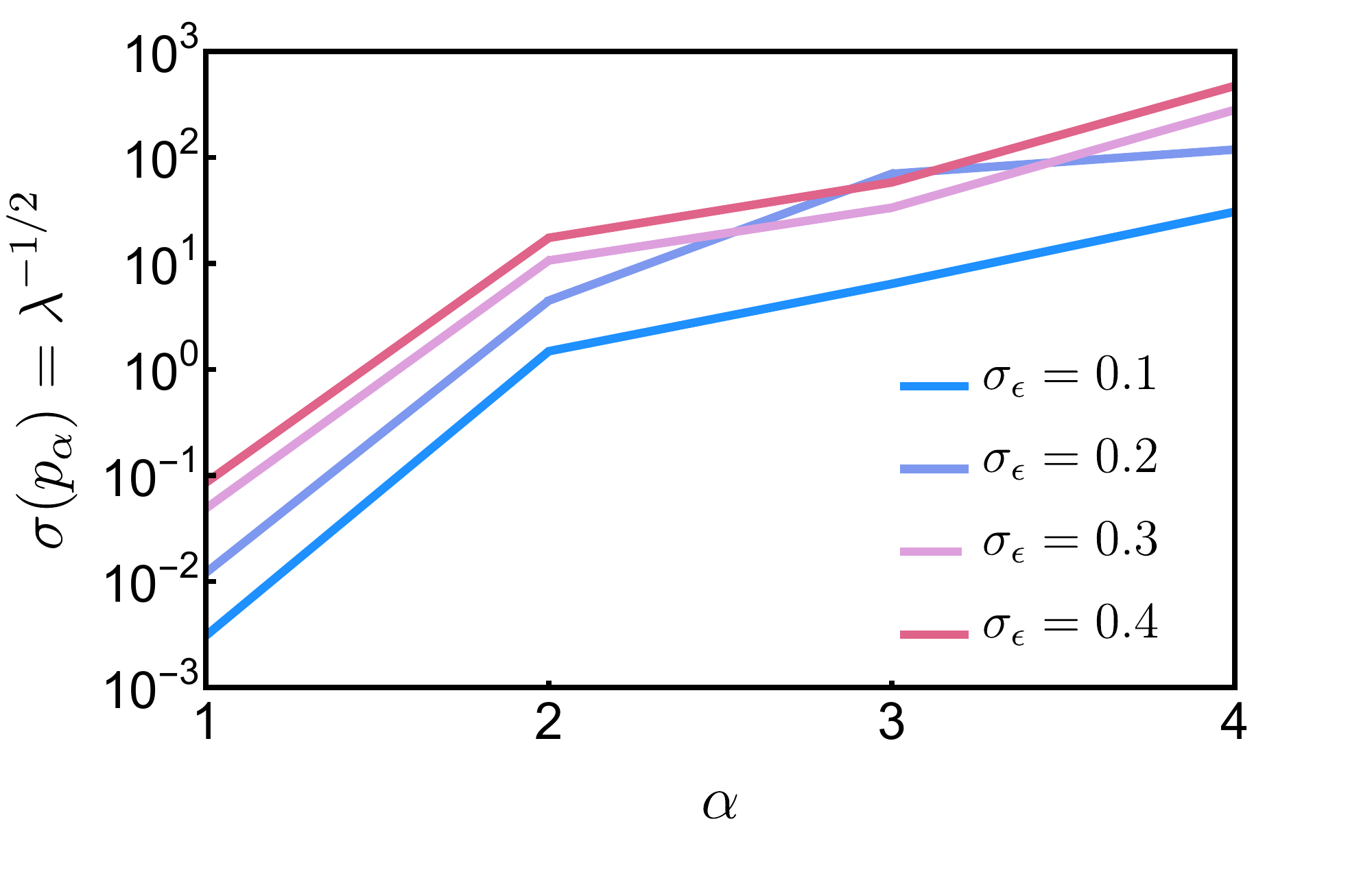}
\includegraphics[width=80mm]{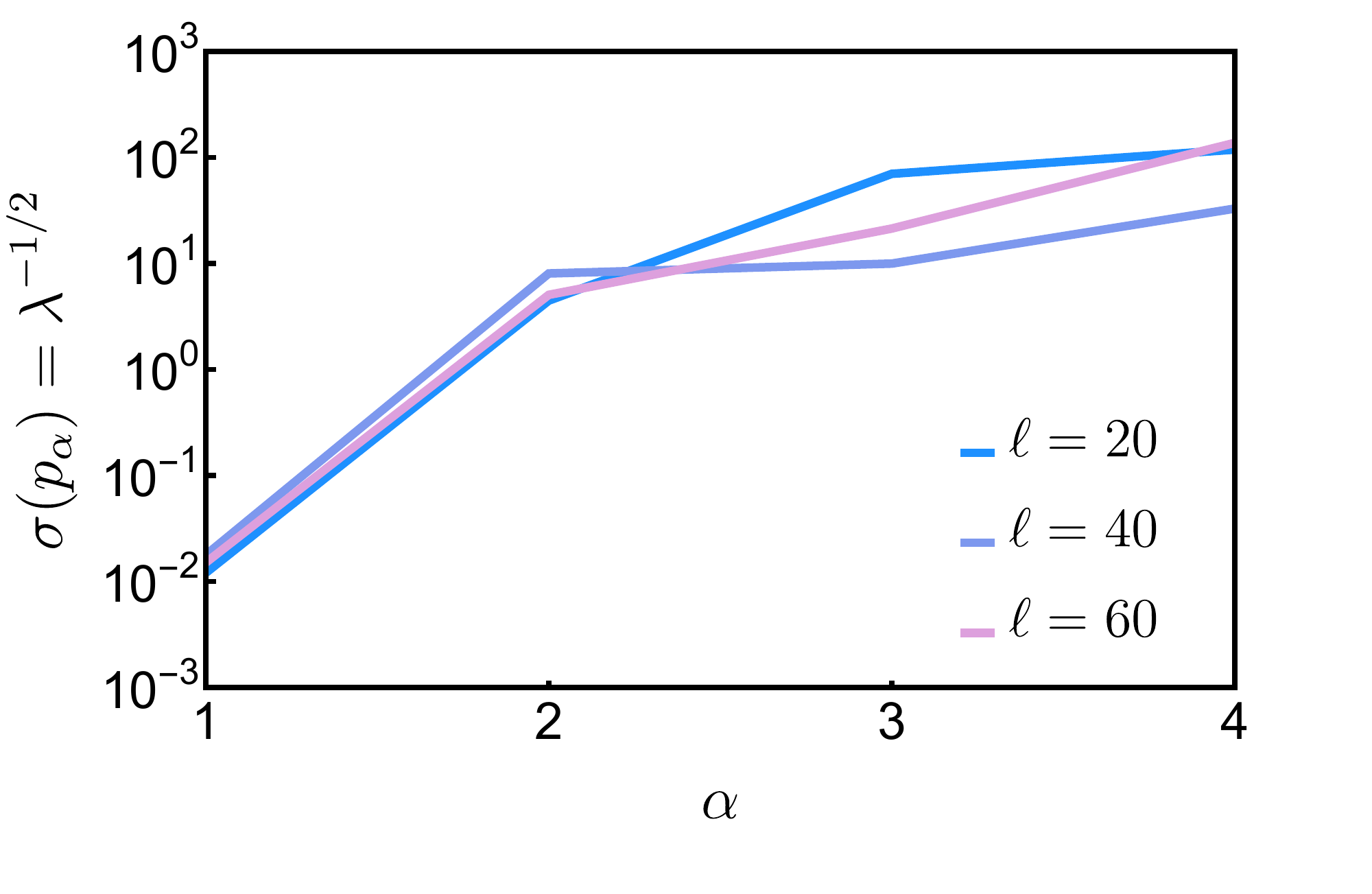}
\includegraphics[width=80mm]{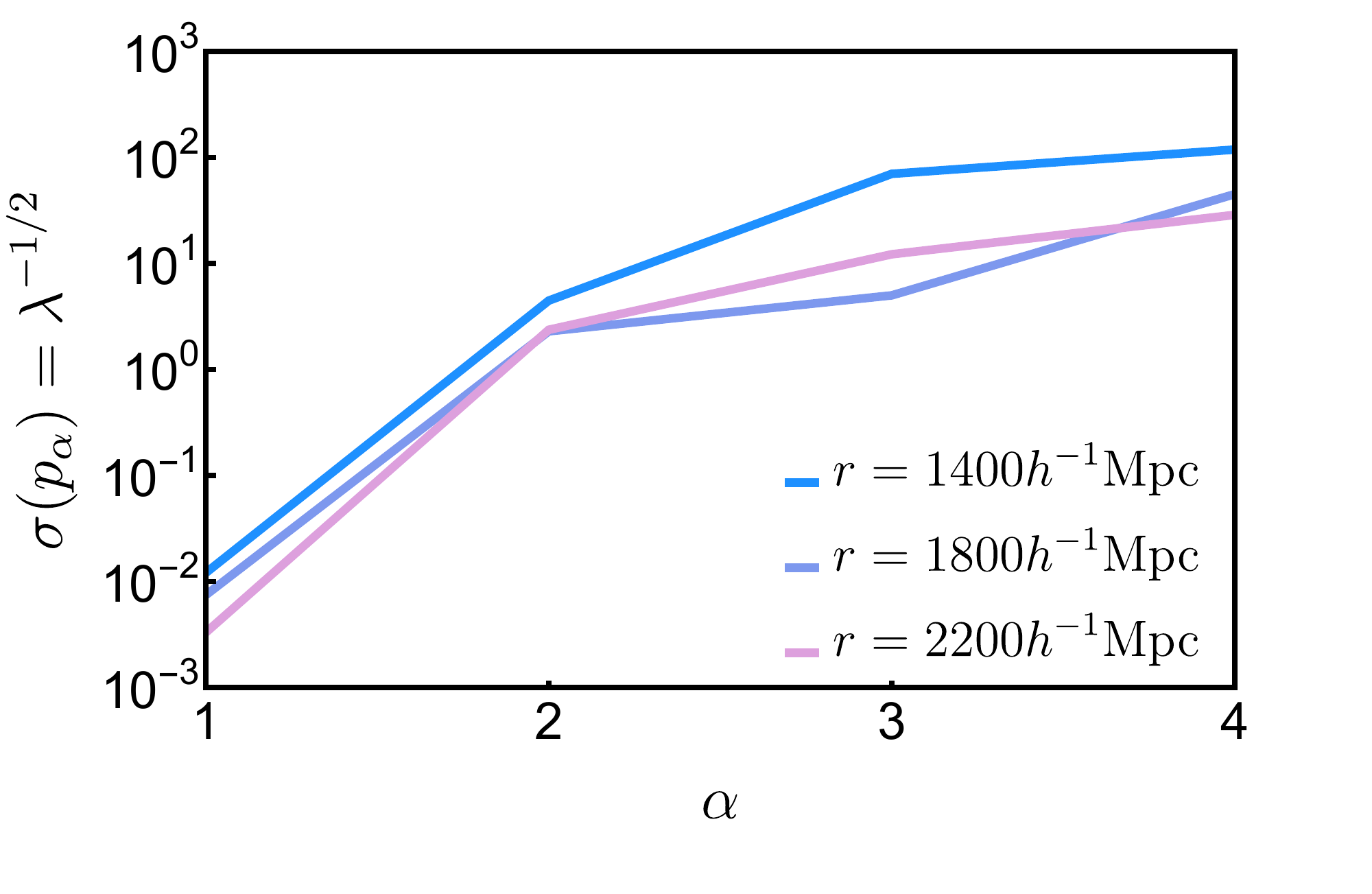}
\caption{
The variance $\sigma (p_{\alpha}) = \lambda^{-1/2}_{\alpha}$ associated with the principal components of the 4x4 Fisher matrix $\bfF$ for the dilaton models. The different curves show the variation of $\sigma (p_{\alpha})$ with increasing ellipticity variance $\sigma_{\epsilon}$. As expected, increasing the noise contribution leads to a greater variance in the eigenvalues. Similarly, increasing the multipole allows us to probe more of the nonlinear features of the power spectra, increasing constraints. However, systematics and un-modelled noise will degrade these values.
}
\label{fig:PCA_dilaton}
\end{center}
\end{figure}
\n
\phantom{TEXT}

\section{Conclusions}
\label{sec:6}

In this paper we have studied the two classes of screened theories of gravity using 3D weak lensing. The focus of this paper has been on understanding the impact of survey configuration and basic noise sources on the constraints that we may place on these theories. The individual spherical Fourier-Bessel spectra can themselves be relatively noisy but by folding in numerous multipoles will place competitive constraints on modified theories of gravity. In Section \ref{sec:4} we outlined the framework used for 3D weak lensing and the construction of the shear lensing spectra. We have not restricted the analysis to survey specific noise curves, which may be dealt with in future studies. We did, however, include ellipticity variance $\sigma^2_{\epsilon}$ in the analysis as this is likely to be a major noise source. Due to the fact that the Weyl lensing potential $\Phi_{\rm{WL}}$ is identical to that predicted in GR, it was shown that the effects in weak lensing from modified theories of gravity only enter the equations through the non-linear matter power spectrum and growth function, simplifying the resulting analysis. 

In order to constrain the model parameters, we implemented a $\chi^2$ analysis for a series of typical surveys with depths $r_0 \in \lbrace 1400, 1800, 2200 \rbrace \, h^{-1} \rm{Mpc}$. In addition, we detailed the impact of varying $\sigma^2_{\epsilon}$ and the multipoles used $\ell$ on the parameter constraints. Increasing the multipole number allows us to probe deeper into the nonlinear regime, where the deviations from GR can become more prominent. However, this also demands that we accurately model the nonlinear matter power spectrum on small scales where systematic and feedback effects, which are poorly understood, will also start to become more prominent. Our approach uses one-loop corrections along with halo modelling in order to extend the domain of validity of the nonlinear matter power spectrum to $k \sim 1 h \, \rm{Mpc}^{-1}$. This places an upper bound on the multipoles used in the analysis. We take an upper limit of $\ell \sim 80$ to our studies in order to restrict ourselves to a regime in which the systematics should be understood and should play a subdominant role.

Our $3 \sigma$ constraints on the $f(R)$ models are $|f_R| < 5 \times 10^{-6}$ for the $n=1$ models and $|f_R| < 9 \times 10^{-6}$ for the $n=2$ models. The Fisher forecasts suggest that these parameters should optimistically be measurable to within a percent level accuracy for upcoming large scale structure surveys. The dilaton constraints at the $3 \sigma$ level are found to be $m0 < 0.4$, $\beta_0 < 0.425$ and $s > 0.37$. The constraints on $r$ are much weaker, with 3D weak lensing seeming to have little sensitivity to the variations in $r$ considered in this paper. The Fisher forecasting typically suggests that the optimistic errors on these parameters are likely to be in the range $10^{-3} - 10^{-1}$ with $r$ again showing much more spread with errors in the region of $\sim$ $\rm{few} \times 10^{-1} - 10^{0}$.  We have not modelled a number of error sources, such as photometric redshift errors, that are likely to degrade our constraints. However, a detailed investigation into optimal binning strategies and optimal weightings should help limit the degradation of parameter constraints in more realistic surveys.  

In this paper we have focussed on the 3D cosmic shear exclusively. A companion paper on testing modified theories of gravity using a 3D analysis of magnification, intrinsic ellipticity distributions and various cross-correlations will be presented elsewhere.  

Finally, we note that the methods presented in this paper may be extended to the K-mouflage models \cite{Babichev09,Brax14}. These models use the non-linear kinetic functions in order to provide a screening mechanism that converges to GR on small astrophysical scales and at high redshifts. In contrast to the $f(R)$ and dilaton models presented in this paper, linear cosmological structures are unscreened and exhibit deviations from $\Lambda$-CDM up to the Hubble scale and the background evolution of dark energy in the K-mouflage models only behaves like a cosmological constant contribution at low redshifts. This can already be seen at the level of the equations of motion in which we find a modified Euler equation, containing an extra friction term and an extra fifth-force potential term, and a modified Poisson equation, containing a time dependent effective Newton constant. These results will be presented in a later paper.              

\section*{Acknowledgments}

G. Pratten acknowledges support from the European Research Council under 
the European Union's Seventh Framework Programme (FP/2007-2013) / ERC Grant Agreement No. [616170].
D. Munshi acknowledges support from the Science and Technology Facilities Council 
(grant number ST/L000652/1).
P. Valageas acknowledges support from the French Agence Nationale de la Recherche under Grant ANR-12-BS05-0002.
P. Brax acknowledges partial support from the European Union FP7 ITN INVISIBLES (Marie Curie Actions, PITN- GA-2011- 289442).

\appendix

\allowdisplaybreaks

\section{Non-Linear Matter Power Spectrum}
\label{app:nlmp}
\subsection{Modified Gravitational Potential}
In this Appendix we schematically outline the formalism used to generate the non-linear matter power spectra. The formalism used is based on standard perturbation theory at one-loop level with a partial resummation of the perturbative series coupled with a halo model \cite{Brax12a,Brax13}. The one-loop corrections allow us to extend the domain of validity of the linear perturbative analysis to $k \leq 0.15 \, h \, \textrm{Mpc}^{-1}$ at the perturbative level while the addition of the halo-model terms further extends the domain of validity to approximately $k \leq 1 \, h \, \textrm{Mpc}^{-1}$, justifying their use in this paper where the scales of most importance are typically less than $\sim 1 h \textrm{Mpc}^{-1}$. 

The scalar-tensor theories in this paper are explicitly coupled to matter via the Jordan-frame metric, which is conformally related to the Einstein-frame metric via $A^2 (\varphi)$ as in Eqn. (\ref{eq:conf}). This leads to an additional fifth force acting on matter particles of mass $m$, ${\bf{F}} = - m c^2 \nabla \, \ln A$. The importance of this term is that it constitutes an additional contribution to the Newtonian term $\Psi_{\rm N}$ in the total gravitational potential
\begin{align}
\label{Psi-tot}
\Psi_{\rm tot} &= \Psi_{\rm N} + \Psi_A .
\end{align}
\n
where we assume that $A(\varphi) \simeq 1$ due to observational bounds. The dynamics of matter particles will implicitly depend on this modified potential. Assuming that the timescale for the evolution of the field perturbations is far below that of cosmological timescale, we can adopt the quasistatic approximation (i.e. the scalar field instantaneously follows the evolution of the matter perturbations) . Care must be taken when using the quasistatic approximation to ensure that we have not introduced significant errors under this assumption. As shown in \cite{Brax13}, the quasistatic approximation will be valid for the scales of interest in 3D weak lensing. The modified gravity potential $\Psi_{\rm tot}$ will be only a functional (i.e. it does not depend on the past history) of the matter density fluctuation $\delta \rho$ in the quasistatic approximation, simplifying the analysis. Solving for $\Psi _{\rm tot}[ \delta \rho ]$, we can then solve for the equations of motion of matter particles in the single-stream approximation (the Euler and continuity equations). 

In a scalar-tensor theory, the modified potential $\Psi_{\rm tot}$ will explicitly depend on the scalar field $\varphi$, demanding that we first solve the Klein-Gordon equation to yield a functional for the field perturbations $\delta \varphi [ \delta \rho ]$. By subtracting the background from the KG equation and expanding in $\delta \varphi$ using the tomographic derivatives in Eqns. (\ref{eqn:beta_n_eqn}-\ref{eqn:kappa_n_eqn}), the KG equation reduces to
\begin{align}
\label{KG-expand}
\left( \frac{\nabla^2}{a^2} - m^2 \right) \cdot \delta \varphi &= \frac{\beta \delta \rho}{c^2 M_{\textrm{pl}}} + \frac{\beta_2 \delta \rho}{c^2 M^2_{\textrm{pl}}} \delta \varphi \\
&\; + \displaystyle\sum^{\infty}_{n=2} \left( \frac{\kappa_{n+1}}{M^{n-1}_{\textrm{pl}}} + \frac{\beta_{n+1} \delta \rho}{c^2 M^{n+1}_{\textrm{pl}}} \right) \frac{\left( \delta \varphi \right)^n}{n !} . \nonumber
\end{align}
\n
This admits a perturbative solution in the nonlinear matter density fluctuations
\begin{align}
\label{phi-rho-hn}
\delta \varphi &= \displaystyle\sum^{\infty}_{n=1} \int d \bfk_1 \dots d \bfk_n \, \delta_D (\bfk_1 + \dots + \bfk_n - \bfk) \\
&\quad \quad \times h_n (\bfk_1 , \dots , \bfk_n ) \, \delta{\rho} (\bfk_1) \dots \delta{\rho} (\bfk_n) \nonumber
\end{align}
\n
where the kernels $h_n$ can be recursively obtained by noting that the left hand side 
in Eqn. (\ref{KG-expand})
is a linear operator that is diagonal in Fourier space and therefore admits an inversion \cite{Brax13}. This expression is then substituted into the expression for the modified gravity potential \ref{Psi-tot}
\begin{align}
\label{Psi-tot-Hn}
\Psi_{\rm tot}(\bfk) &= \displaystyle\sum_{n=1}^{\infty} \int d \bfk_1 \dots d \bfk_n \, \delta_D (\bfk_1 + \dots \bfk_n - \bfk ) \\
&\qquad \times H_n(\bfk_1 , \dots , \bfk_n ) \, \delta{\rho} (\bfk_1) \dots \delta{\rho}(\bfk_n) .\nonumber
\end{align}
\n
This method directly applies to the Dilaton models introduced in 
section~\ref{Dilaton-Models}, where the scalar field $\varphi$ obeys the Klein-Gordon,
Eqn. (\ref{KG-Dilatons}), and the coefficients $\beta_n$ and $\kappa_n$ of
\ref{KG-expand} were defined in Eqns. (\ref{eqn:beta_n_eqn}-\ref{eqn:kappa_n_eqn}).
For the $f(R)$ theories introduced in section~\ref{fR-models}, 
instead of the fluctuations of the scalar field $\delta\varphi$ we need to solve
for the fluctuations of the Ricci scalar $\delta R$. From the constraint equation
\ref{deltaR-deltarho} we obtain the functional $\delta R[\delta\rho]$, as a perturbative
expansion over $\delta\rho$ in a fashion similar to \ref{phi-rho-hn}.
Using Eqn. (\ref{eqn:fRPhi}) this provides, in turn, the perturbative expansion of the total potential 
as in \ref{Psi-tot-Hn}.

In the case of GR, the Poisson equation is linear and only the first kernel $H_1$ is non-zero.
For the modified-gravity scenarios, the expansion \ref{Psi-tot-Hn} exhibits terms at all orders
because the non-linear Klein-Gordon equation \ref{KG-expand} generates terms
of all orders for the functionial $\delta\varphi[\delta\rho]$, see \ref{phi-rho-hn}, and the
non-linear fifth-force potential $A(\varphi)$ generates further non-linear terms for
$A[\delta\rho] \equiv A[ \delta\varphi [ \delta\rho ] ]$.

\subsection{Single Stream Approximation}
The next key step is to propagate the modified gravitational potential through the hydrodynamical equations of motion in the perturbative regime. Using the single-stream approximation \cite{Brax13,Brax14}, which is valid on large scales, the dynamics of the matter fluid is given by the continuity and Euler equations
\begin{align}
\frac{\partial \delta}{\partial \tau} + \nabla \cdot \left[ (1 + \delta ) \, {\bf{v}} \right] &= 0 , \\
\frac{\partial {\bf{v}}}{\partial \tau} + \mathcal{H} \, {\bf{v}} + \left( {\bf{v}} \cdot \nabla \right) \, {\bf{v}} &= - \nabla \cdot \Psi_{\rm tot} ,
\end{align}
\n
where $\mathcal{H}$ the conformal Hubble expansion rate and ${\bf{v}}$ the peculiar
velocity. Introducing a time variable $\eta = \ln a$ and a two-component vector 
\cite{Crocce05}
\begin{align}
\psi &= \begin{bmatrix}
           \psi_1 \\
           \psi_2
         \end{bmatrix}
         =
         \begin{bmatrix}
           \delta \\
           - (\nabla \cdot {\bf{v}}) / \mathcal{H}
         \end{bmatrix}
\end{align}
\n
the equations of motion can be greatly simplified in the Fourier domain to \cite{Brax13}
\begin{align}
\label{eqn:SSFourier1}
&\frac{\partial {\psi}_1}{\partial \eta} - {\psi_2} = \int d \bfk_1 \, d \bfk_2 \, \delta_D (\bfk_1 + \bfk_2 - \bfk) \, \\
&\qquad \qquad \nonumber \times \hat{\alpha} (\bfk_1 , \bfk_2 ) \, {\psi}_2 (\bfk_1) \, {\psi}_1 (\bfk_2) ,\\
\label{eqn:SSFourier2}
&\frac{\partial {\psi}_2}{\partial \eta} + \frac{k^2}{a^2 H^2} {\Psi}_{\rm tot} + \frac{1+ 3 \, \Omega_{\Lambda}}{2} {\psi}_2 = \int d \bfk_1 \, d \bfk_2 \, \\
&\qquad \nonumber \times \delta_D (\bfk_1 + \bfk_2 - \bfk) \hat{\beta} (\bfk_1 , \bfk_2) \, {\psi}_2 (\bfk_1) \, {\psi}_2 (\bfk_2) ,
\end{align}
\n
where the coupling kernels are explicitly given by \cite{Brax13}
\begin{align}
\hat{\alpha} (\bfk_1 , \bfk_2) &= \frac{(\bfk_1 + \bfk_2) \cdot \bfk_1}{k^2_1} , \\
\hat{\beta} (\bfk_1 , \bfk_2) &= \frac{| \bfk_1 + \bfk_2 |^2 \, (\bfk_1 \cdot \bfk_2 )}{2 k^2_1 k^2_2} .
\end{align}
\n
In GR, the Newtonian gravitational potential is linear in the density field and therefore the Euler and continuity equations are quadratic. As detailed in \cite{Brax12a,Brax13}, however, in the modified theories of gravity considered in this paper, the potential $\Psi_{\rm tot}$ 
of Eq. (\ref{Psi-tot-Hn}) is nonlinear and contains terms at all orders in $\delta \rho$. This necessitates that we include vertices to all orders. Schematically, this can be done by re-expressing Eqns. (\ref{eqn:SSFourier1}-\ref{eqn:SSFourier2}) as 
\begin{align}
\mathcal{O} (x ,x^{\prime} ) \cdot {\psi} (x^{\prime}) &= \displaystyle\sum^{\infty}_{n=2} \, K^s_n (x ; x_1 , \dots x_n ) \cdot {\psi} (x_1) \dots {\psi} (x_n) ,
\label{eqn:EOMallvert}
\end{align}
\n
where $x = (\bfk , \eta , i)$ and $i \in \lbrace 1,2 \rbrace$ denotes the index of ${\psi}$. The matrix $\mathcal{O}$ is written in terms of a function $\epsilon(k,\eta)$ that measures the deviation from the Newtonian gravitational potential at linear order
\begin{align}
\mathcal{O} (x , x^{\prime}) &= \delta_D (\eta^{\prime} - \eta) \delta_D (\bfk^{\prime} - \bfk) \\
&\quad \times \nonumber
\begin{bmatrix}
\frac{\partial}{\partial \eta} & -1 \\
-\frac{3}{2} \Omega_m (\eta) (1 + \epsilon (k,\eta)) & \frac{\partial}{\partial \eta} + \frac{1+3 \Omega_{\Lambda} (\eta)}{2} 
\end{bmatrix} .
\end{align}
\n
The vertices $K^s_n$ are equal-time vertices with the schematic form
\begin{align}
K^s_n (x, x_1 , \dots , x_n ) &= \delta_D (\eta_1 - \eta) \dots \delta_D (\eta_n -\eta) \\
&\quad \times \nonumber \delta_D (\bfk_1 + \dots + \bfk_n - \bfk) \\
&\quad \times \nonumber \gamma^s_{i;i_1 \dots i_n} (\bfk_1 , \dots , \bfk_n ; \eta) .
\end{align}
\n
The vertices $\gamma^s_{i;i_1 \dots i_n}$ for GR, $f(R)$ and the scalar-tensor theories may be found in \cite{Brax12a,Brax13,Brax14}.

\subsection{One-Loop Matter Power Spectrum}
Given the equation of motion incorporating vertices at all orders in Eqn. (\ref{eqn:EOMallvert}), it is possible to calculate the power spectrum up to the required order in perturbation theory. This paper uses power spectra that only incorporate the one-loop diagrams, corresponding to corrections that are third order in the fields \cite{Brax12a,Brax13,Brax14}. This proceeds by looking for a solution of the nonlinear equation of motion as a perturbative expansion in powers of the linear growing mode $\psi_L$
\begin{align}
\psi(x) &= \displaystyle\sum^{\infty}_{n=1} \psi^{(n)} (x) \quad \textrm{such that} \quad \psi^{(n)} \propto \psi^n_L .
\end{align}
\n
The linear order equation of motion is simply $\mathcal{O} \cdot {\psi}_L = 0$, leading to the usual two growing modes $D_{\pm} (k , \eta)$ with the concomitant evolution equation
\begin{align}
\frac{\partial^2 D}{\partial \eta^2} + \frac{1 + 3 \Omega_{\Lambda}}{2} \frac{\partial D}{\partial \eta} - \frac{3}{2} \Omega_M (1 + \epsilon) D = 0,
\end{align}
\n
where the initial conditions are set by 
\begin{align}
t \rightarrow 0 : \quad D_{+} \rightarrow a = e^{\eta} , \quad D_{-} \propto a^{-3/2} = e^{-3 \eta / 2} .
\end{align}
\n
In GR, the modes $D_{\pm}$ are $k$-independent. However, in modified theories of gravity, the modes will be $k$-dependent due to the presence of the new $\epsilon (k ,\eta)$ term. As the decaying modes typically become negligible, the first-order solution simplifies nicely to
\begin{align}
{\psi}^{(1)} &= {\psi}_L = {\delta}_{L0} (\bfk) 
\begin{bmatrix}
D_{+} (k ,\eta) \\
\frac{\partial D_{+}}{\partial \eta} (k,\eta) 
\end{bmatrix} ,
\end{align}
\n
with the linear density field $\delta_{L0} (\bfk)$ fully determining the initial conditions.

\begin{figure}[t!]
\begin{center}
\includegraphics[width=85mm]{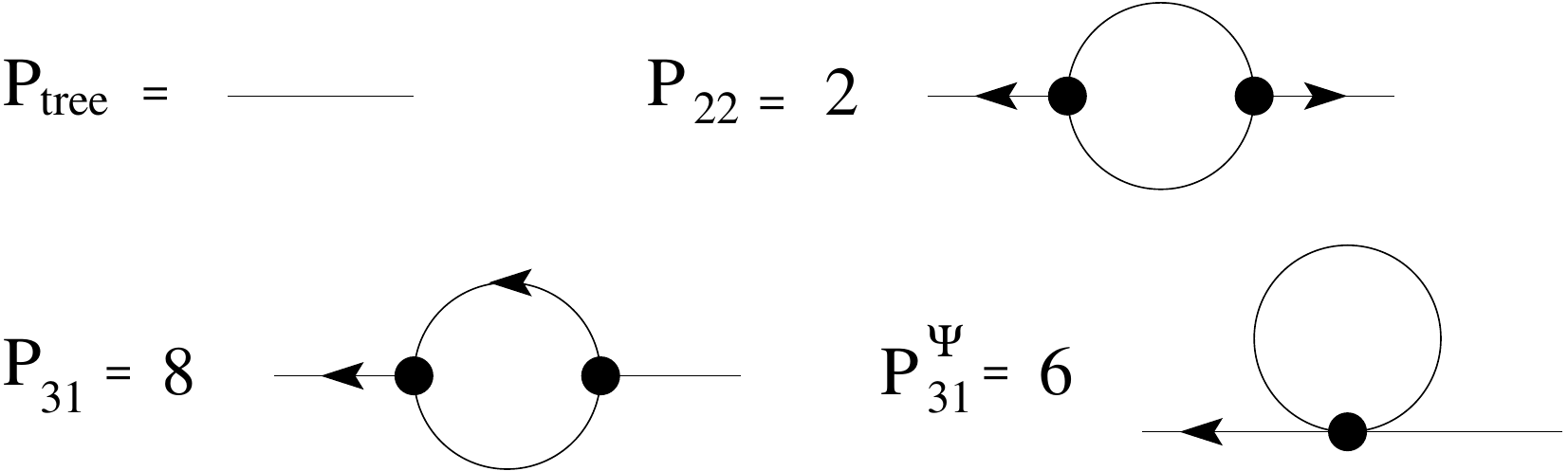}
\caption{
Contributions at order $P^2_L$ to the matter power spectrum from loop-diagrams. Schematically we find $P(k) = P_{\rm{tree}} (k) + P_{22} (k) + P_{31} (k) + P^{\Psi}_{31} (k)$. The black dots are vertices for $K^s_n$ and the lines with an arrow denote the retarded propagator $R_L$. Lines without an arrow correspond to the linear correlator $C_L$. Figure taken from \cite{Brax13}.
}
\label{fig:P1loop}
\end{center}
\end{figure}

The higher order terms $\psi^{(n)}$ are obtained via recursion of Eqn. (\ref{eqn:EOMallvert}) in terms of the retarded Green's function $R_L$ \cite{Brax13}
\begin{align}
\mathcal{O}(x,x^{\prime}) \cdot R_L (x^{\prime} , x^{\prime \prime}) &= \delta_D (x - x^{\prime \prime}),
\end{align}
\n
implicitly demanding that
\begin{align}
R_L (x_1 , x_2) = 0 \quad \textrm{for} \quad \eta_1 < \eta_2 .
\end{align}
\n
This leads to the second and third order fields
\begin{align}
{\psi}^{(2)} &= R_L \cdot K^s_2 \cdot {\psi}^{(1)} {\psi}^{(1)} , \\
{\psi}^{(3)} &= 2 R_L \cdot K^s_2 \cdot {\psi}^{(2)} {\psi}^{(1)} + R_L \cdot K^s_3 \cdot {\psi}^{(1)} {\psi}^{(1)} {\psi}^{(1)} .
\end{align}
\n
Up to order $\psi^{(4)}_L$, the 2-point correlation function can be written as 
\begin{align}
C_2 (x_1 , x_2) &= \langle {\psi} (x_1) {\psi} (x_2) \rangle \\
&= \langle {\psi}^{(1)} {\psi}^{(1)} \rangle + \langle {\psi}^{(2)} {\psi}^{(2)} \rangle + \langle {\psi}^{(3)} {\psi}^{(1)} \rangle \\
&\qquad + \langle {\psi}^{(1)} {\psi}^{(3)} \rangle + \mathcal{O} (\psi^{(6)}_L) . \nonumber
\end{align}
\n
Substituting the expressions for ${\psi}^{(1)}$, ${\psi}^{(2)}$ and ${\psi}^{(3)}$ into the 2-point correlation function and defining the equal-time matter density power spectrum
\begin{align}
\langle {\delta} (\bfk_1 , \eta) {\delta} (\bfk_2 ,\eta) \rangle &= \delta_D (\bfk_1 + \bfk_2) \, P(k_1 , \eta) ,
\end{align}
\n
we can obtain the power spectrum up to order $P^2_L$ via Wick's theorem
\begin{align}
P(k) &= P_{\rm{tree}} (k) + P_{\rm{1-loop}} (k) \\
P_{\rm{tree}} &= P_L (k) \\
P_{\rm{1-loop}} &= P_{22} + P_{31} + P^{\Psi}_{31} ,
\end{align}
\n
where $P_{31}$ terms encapsulate contributions from the 31 and 13 terms above. Both $P_{22}$ and $P_{31}$ are present in GR whereas $P^{\Psi}_{31}$  is a genuinely new effect generated by the modified theory of gravity \cite{Brax13}. The 1-loop diagrams that contribute to the power spectrum are explicitly shown in Figure \ref{fig:P1loop}. Note that the contributions to $P_{31}$ in GR will have different vertices and linear propagators to modified theories of gravity. In particular, the linear propagator will become momenta dependent and the vertices time dependent. Explicit expressions for $P_{31}$ and $P^{\Psi}_{31}$ can be found in \cite{Brax13} as momenta integrals over the propagators $R_L$, the vertices $\gamma^s$ and the correlators $C_L$

\subsection{Halo Model Matter Power Spectrum}
Standard 1-loop perturbation theory breaks down relatively quick as we enter the nonlinear regime. In order to extend the domain of validity of this perturbative approach, we can combine perturbation theory with halo models to generate a matter power spectrum down to the smaller, highly nonlinear scales. The halo model schematically provides a  matter power spectrum of the form 
\begin{align}
P(k) &= P_{\rm{1h}} (k) + P_{\rm{2h}} (k) ,
\end{align}
\n
where $P_{\rm{1h}}$ models the contribution to the matter power from pairs within the same halo and $P_{\rm{2h}}$ models contributions from pairs in two separate halos. The key quantities that enter the halo model are the normalised halo mass function $f(\nu)$ and a variance weighted density threshold $\nu$ defined by
\begin{align}
n(M) \frac{d M}{M} &= \frac{\bar{\rho}}{M} f(\nu) \frac{d \nu}{\nu} , \quad \textrm{where} \quad \nu = \frac{\delta_L (M)}{\sigma_L (M)} .
\end{align}
\n
The 1-halo contribution is given by \cite{Brax13}
\begin{align}
P_{\rm{1h}} (k) &= \int^{\infty}_0 \frac{d \nu}{(2 \pi)^3} \, f(\nu) \, \frac{M}{\bar{\rho} \nu} \, \left[ {u}_M (k) - {W}(k q_M) \right] ,
\end{align}
\n
with ${u}_M(k)$ the normalised Fourier transform of the halo radial profile and ${W} (k q_M)$ the normalised Fourier transform of the top hat of Lagrangian radius $q_M$. The 2-halo term is defined by \cite{Brax13}
\begin{align}
P_{\rm{2h}} (k) &= \int \frac{d \Delta \bfq}{(2 \pi)^3} \, F_{2h} (\Delta q) \, \langle e^{i \bfk \cdot \Delta \bfx} \rangle^{\rm{vir}}_{\Delta q} \frac{1}{1+ A_1} \\
&\quad \times  e^{- \frac{1}{2} k^2 (1 - \mu^2) \sigma^2_{\perp}} \Bigg\lbrace e^{- \varphi_{\parallel} (- i k \mu \Delta q \sigma^2_{k_{\parallel}})} + A_1 \nonumber \\
&\quad + \displaystyle\int\limits^{0^{+} + i \infty}_{0^{+} - i \infty} \frac{d y}{2 \pi i} e^{- \varphi_{\parallel} (y) / \sigma^2_{k_{\parallel}}} \left( \frac{1}{y} - \frac{1}{y + i k \mu \Delta q \sigma^2_{k_{\parallel}}} \right) \nonumber
\Bigg\rbrace .
\end{align}
\n
This is a rather complicated expression and explicit details on its derivation can be found in \cite{Brax13}. Here, we just note that this expression relates the power spectrum to the statistics of the Eulerian separation, $\Delta \bfx = \bfx_2 - \bfx_1$, of pairs of particles with initial Lagrangian separation $\Delta \bfq = \bfq_2 - \bfq_1$. The factor $F_{2h}$ encapsulates the probability that a pair of particles of separation $\Delta \bfq$ are in separate halos. The exponential average $\langle e^{i \bfk \cdot \Delta \bfx} \rangle^{\rm{vir}}_{\Delta q}$ is the contribution due to internal motions within each virialised halo. The Gaussian pre-factor $e^{- \frac{1}{2} k^2 (1 - \mu^2) \sigma^2_{\perp}}$ describes the contribution from large-scale longitudinal motions with $\sigma^2_{k_{\parallel}}$ the variance of the longitudinal relative displacement. The factor $A_1$ and the complex integral arise from adhesion-like regularisation that aims to capture the formation of pancakes. 

\section{Principal Component Analysis: Eigenvalues of the Fisher Matrix}
\label{App:PCA}
Here we give an example of the Fisher matrix, decorrelation matrix and eigenvalues for the $\ell = 20$ and $\sigma_{\epsilon} = 0.2$ configuration. The Fisher matrix $F_{\alpha \beta}$ is given in Table \ref{tabular:tabFab} and the decorrelation matrix $W_{\alpha \beta}$ by Table \ref{tabular:tabWab}. The Fisher matrix is related to the decorrelation matrix via ${\bf{F}} = {\bf{W}}^T \Lambda {\bf{W}}$, where $\Lambda$ is the diagonal matrix constructed from the eigenvalues $\lambda_{\alpha}$
\begin{align}
\nonumber
\lambda_1 &= 6.869 \times 10^{+3},  \; \lambda_2 = 5.014 \times 10^{-2}, \\ 
\lambda_3 &= 2.036 \times 10^{-4}, \; \lambda_4 = 7.104 \times 10^{-5} .
\end{align} 
\n
The condition number of the matrix $\kappa$ is particularly large at $9.67 \times 10^9$. The $1\sigma$ MVB is reconstructed by a weighted sum over the eigenvalues as per Eqn. (\ref{eqn:MVB})
\begin{align}
\Delta \theta_{\alpha} &= \left( \displaystyle\sum_{\beta = 1}^{n} W^2_{\alpha \beta} \lambda^{-1}_{\beta} \right)^{1/2}.
\end{align}
\n
We can check how $\Delta \theta_{\alpha}$ varies as a function of the number of eigenvalues included in the analysis. This is shown in Figure \ref{fig:MVBEVs}.

{\renewcommand{\arraystretch}{1.3}%
\begin{table}[h!]
\begin{center}
\begin{tabular}{| c | c | c | c | c | }
  \hline
  $F_{\alpha \beta}$ & $m_0$ & $r$ & $\beta_0$  & $s$ \\
  \hline
    \rowcolor[gray]{0.8} 
  $m_0$  & $+ 6.062 \times 10^3$ & $-1.739 \times 10^2$ & $-1.969 \times 10^3$ & $+ 9.926 \times 10^2$ \\
  \hline
  $r$  & $-1.739 \times 10^2$ & $+ 4.991 \times 10^0$ & $+ 5.647 \times 10^1$ & $-2.848 \times 10^1$ \\
  \hline
   \rowcolor[gray]{0.8} 
  $\beta_0$  & $-1.969 \times 10^3$ & $+ 5.647 \times 10^1$ & $+ 6.393 \times 10^2$ & $-3.224 \times 10^2$ \\
  \hline
  $s$  & $+ 9.926 \times 10^2$ & $-2.848 \times 10^1$ & $-3.224 \times 10^2$ & $+ 1.626 \times 10^2$ \\
  \hline
\end{tabular}
\caption{Fisher matrix $F_{\alpha \beta}$ for the dilaton parameters $\lbrace m_0 , r , \beta_0 , s \rbrace$.}
\label{tabular:tabFab}
\end{center}
\end{table}

{\renewcommand{\arraystretch}{1.3}%
\begin{table}[h!]
\begin{center}
\begin{tabular}{| c | c | c | c | c | }
  \hline
  $W_{\alpha \beta}$ & $\beta=1$ & $2$ & $3$ & $4$ \\
  \hline
    \rowcolor[gray]{0.8} 
  $\alpha = 1$       & $-0.9394$   & $+0.02694$ & $+0.3051$ & $-0.1538$ \\
  \hline
  $2$                & $-0.3330$   & $-0.3072$  & $-0.7894$ & $+0.4141$ \\
  \hline
   \rowcolor[gray]{0.8} 
  $3$                & $-0.05843$  & $+0.6421$  & $-0.5194$ & $-0.5608$ \\
  \hline
  $4$                & $-0.00562$  & $+0.7019$  & $+0.1179$ & $+0.7002$ \\
  \hline
\end{tabular}
\caption{Decorrelation matrix $W_{\alpha \beta}$ for the dilaton Fisher matrix $F_{\alpha \beta}$.}
\label{tabular:tabWab}
\end{center}
\end{table}

\begin{figure}[t!]
\begin{center}
\includegraphics[width=80mm]{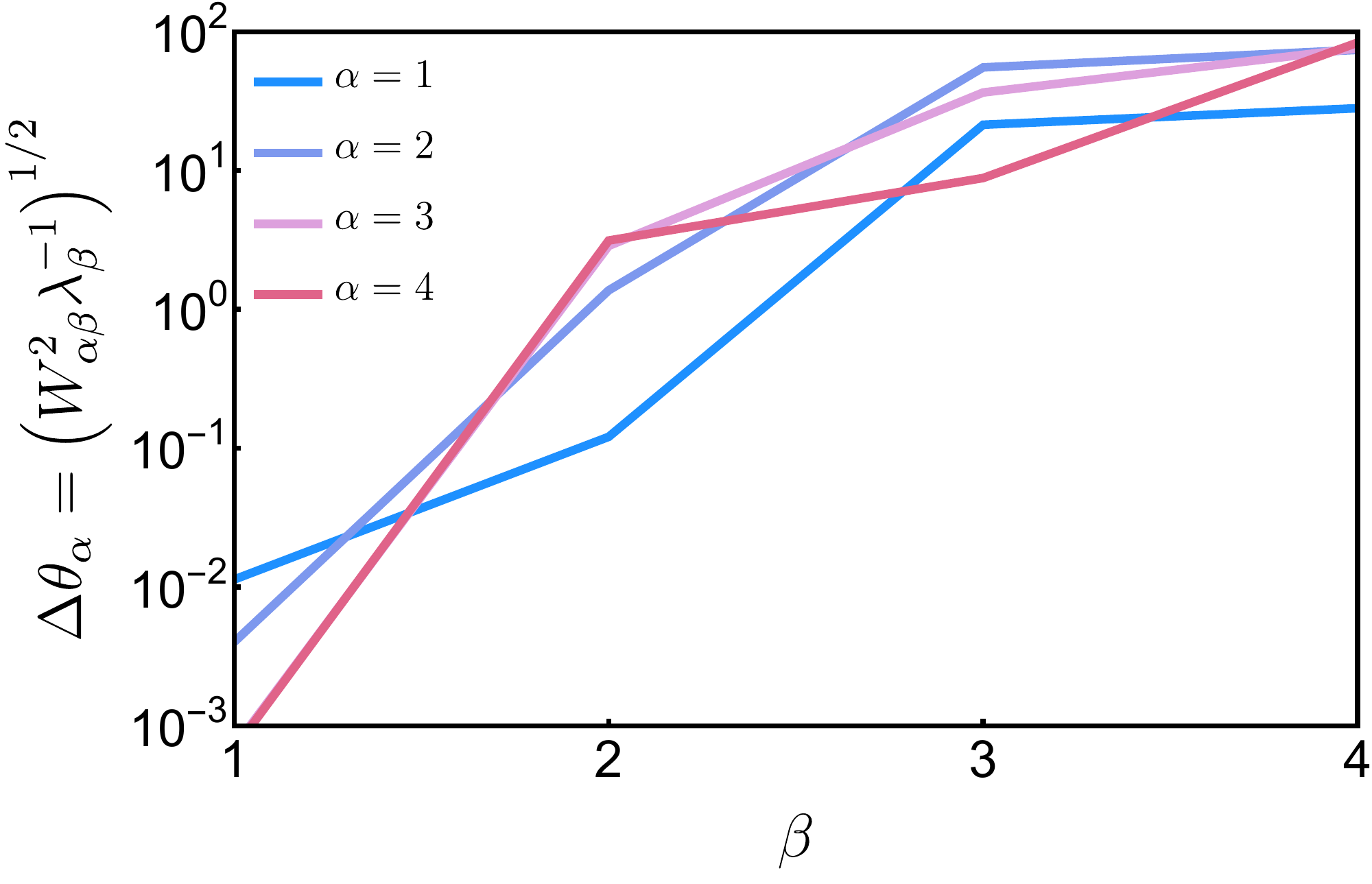}
\caption{
The minimum variance bound $\Delta \theta_{\alpha}$ as a function of the eigenvalues $\lambda_{\beta}$ included, as per Eqn. (\ref{eqn:MVB}).
}
\label{fig:MVBEVs}
\end{center}
\end{figure}

\newpage



\bibliography{3D_WL_MG}

\end{document}